\PassOptionsToPackage{unicode}{hyperref}
\PassOptionsToPackage{hyphens}{url}
\PassOptionsToPackage{dvipsnames,svgnames,x11names}{xcolor}
\documentclass[
  12pt]{article}

\usepackage{amsmath,amssymb,amsthm}
\usepackage{iftex}
\ifPDFTeX
  \usepackage[T1]{fontenc}
  \usepackage[utf8]{inputenc}
  \usepackage{textcomp} 
\else 
  \usepackage{unicode-math}
  \defaultfontfeatures{Scale=MatchLowercase}
  \defaultfontfeatures[\rmfamily]{Ligatures=TeX,Scale=1}
\fi
\usepackage{lmodern}
\ifPDFTeX\else  
\fi
\IfFileExists{upquote.sty}{\usepackage{upquote}}{}
\IfFileExists{microtype.sty}{
  \usepackage[]{microtype}
  \UseMicrotypeSet[protrusion]{basicmath} 
}{}
\makeatletter
\@ifundefined{KOMAClassName}{
  \IfFileExists{parskip.sty}{%
    \usepackage{parskip}
  }{
    \setlength{\parindent}{0pt}
    \setlength{\parskip}{6pt plus 2pt minus 1pt}}
}{
  \KOMAoptions{parskip=half}}
\makeatother
\usepackage{xcolor}
\makeatletter
\ifx\paragraph\undefined\else
  \let\oldparagraph\paragraph
  \renewcommand{\paragraph}{
    \@ifstar
      \xxxParagraphStar
      \xxxParagraphNoStar
  }
  \newcommand{\xxxParagraphStar}[1]{\oldparagraph*{#1}\mbox{}}
  \newcommand{\xxxParagraphNoStar}[1]{\oldparagraph{#1}\mbox{}}
\fi
\ifx\subparagraph\undefined\else
  \let\oldsubparagraph\subparagraph
  \renewcommand{\subparagraph}{
    \@ifstar
      \xxxSubParagraphStar
      \xxxSubParagraphNoStar
  }
  \newcommand{\xxxSubParagraphStar}[1]{\oldsubparagraph*{#1}\mbox{}}
  \newcommand{\xxxSubParagraphNoStar}[1]{\oldsubparagraph{#1}\mbox{}}
\fi
\makeatother

\usepackage{longtable,booktabs,array}
\usepackage{calc} 
\usepackage{etoolbox}
\makeatletter
\patchcmd\longtable{\par}{\if@noskipsec\mbox{}\fi\par}{}{}
\makeatother
\IfFileExists{footnotehyper.sty}{\usepackage{footnotehyper}}{\usepackage{footnote}}
\makesavenoteenv{longtable}
\usepackage{graphicx}
\makeatletter
\def\maxwidth{\ifdim\Gin@nat@width>\linewidth\linewidth\else\Gin@nat@width\fi}
\def\maxheight{\ifdim\Gin@nat@height>\textheight\textheight\else\Gin@nat@height\fi}
\makeatother
\setkeys{Gin}{width=\maxwidth,height=\maxheight,keepaspectratio}
\makeatletter
\def\fps@figure{htbp}
\makeatother

\makeatletter
\@ifpackageloaded{caption}{}{\usepackage{caption}}
\AtBeginDocument{%
\ifdefined\contentsname
  \renewcommand*\contentsname{Table of contents}
\else
  \newcommand\contentsname{Table of contents}
\fi
\ifdefined\listfigurename
  \renewcommand*\listfigurename{List of Figures}
\else
  \newcommand\listfigurename{List of Figures}
\fi
\ifdefined\listtablename
  \renewcommand*\listtablename{List of Tables}
\else
  \newcommand\listtablename{List of Tables}
\fi
\ifdefined\figurename
  \renewcommand*\figurename{Figure}
\else
  \newcommand\figurename{Figure}
\fi
\ifdefined\tablename
  \renewcommand*\tablename{Table}
\else
  \newcommand\tablename{Table}
\fi
}
\@ifpackageloaded{float}{}{\usepackage{float}}
\floatstyle{ruled}
\@ifundefined{c@chapter}{\newfloat{codelisting}{h}{lop}}{\newfloat{codelisting}{h}{lop}[chapter]}
\floatname{codelisting}{Listing}

\makeatother
\makeatletter
\@ifpackageloaded{caption}{}{\usepackage{caption}}
\@ifpackageloaded{subcaption}{}{\usepackage{subcaption}}
\makeatother

\ifLuaTeX
  \usepackage{selnolig}  
\fi
\usepackage[]{natbib}
\usepackage{bookmark}

\IfFileExists{xurl.sty}{\usepackage{xurl}}{} 
\hypersetup{
  pdftitle={Exact Likelihood and Sampling for Riemannian Gaussian Distributions on Correlation Matrices},
  pdfauthor={Kisung You},
  pdfkeywords={affine-invariant metric; quotient manifold; normalizing constant; Frechet estimation; Fisher transformation; Monte Carlo integration},
  colorlinks=true,
  linkcolor={blue},
  filecolor={Maroon},
  citecolor={Blue},
  urlcolor={Blue},
  pdfcreator={LaTeX via pandoc}}

\usepackage{algorithm}
\usepackage{algpseudocode}
\usepackage{enumitem}
\usepackage{tikz} 
\usetikzlibrary{arrows.meta,positioning}
\usepackage{xspace}
\newcommand{\Frechet}{Fr\'{e}chet\xspace}
\newcommand{\SPD}{\mathcal S_{++}}
\newcommand{\Sym}{\mathbb S}
\newcommand{\Corr}{\mathcal C^+}
\newcommand{\Diagp}{\operatorname{Diag}_{+}}
\newcommand{\diag}{\operatorname{diag}}
\newcommand{\Diag}{\operatorname{Diag}}
\newcommand{\tr}{\operatorname{tr}}
\newcommand{\vecl}{\operatorname{vecl}}
\newcommand{\vol}{\operatorname{vol}}
\newcommand{\Log}{\operatorname{Log}}
\newcommand{\Exp}{\operatorname{Exp}}
\newcommand{\Scal}{\operatorname{Scal}}
\newcommand{\Ric}{\operatorname{Ric}}
\newcommand{\E}{\mathbb E}
\newcommand{\Var}{\operatorname{Var}}
\newcommand{\R}{\mathbb R}
\newcommand{\QA}{\mathrm{QA}}
\newcommand{\AI}{\mathrm{AI}}
\newcommand{\NR}{\mathcal N^{\mathrm R}_{\mathrm{QA}}}

\newcommand{\argmin}{\operatorname*{arg\,min}}
\newcommand{\Had}{\circ}

\newcommand{\ess}{\operatorname{ESS}}

\newtheorem{theorem}{Theorem}
\newtheorem{proposition}{Proposition}
\newtheorem{lemma}{Lemma}
\newtheorem{corollary}{Corollary}

\newtheorem{definition}{Definition}

\newcommand{\anon}{1}

\begin{document}

\def\spacingset#1{\renewcommand{\baselinestretch}%
{#1}\small\normalsize} \spacingset{1}


\if1\anon
{
  \title{\bf Exact Likelihood and Sampling for Riemannian Gaussian Distributions on Correlation Matrices}
    \author{Kisung You}
\date{}
  \maketitle
} \fi

\if0\anon
{
  \bigskip
  \bigskip
  \bigskip
  \begin{center}
    {\LARGE\bf Title}
\end{center}
  \medskip
} \fi

\bigskip
\begin{abstract}
Correlation matrices arise when marginal scales are removed from covariance matrices, yet a normalized likelihood must account for both quotient distance and quotient volume. We propose a Riemannian Gaussian model for full-rank correlation matrices under quotient-affine geometry. The distribution is proper and has finite radial moments. We derive exact score and profiled-scale equations and recover Fisher-transformed Gaussian inference for two-dimensional matrices. In higher dimension, a curvature calculation shows that the normalizing constant can vary with the center. Exact maximum likelihood and \Frechet estimation may therefore have different population targets. We develop chart-based methods for evaluating the normalizer, fitting the likelihood, and sampling. Numerical studies verify the analytic case and compare integration, estimation, and sampling procedures across dimensions and dispersion regimes. A rolling-finance application and a controlled prior study illustrate both the value and computational cost of the model. The method is most reliable in small to moderate dimensions, while proposal efficiency and numerical conditioning deteriorate near the boundary and at larger dispersion.
\end{abstract}

\noindent%
\vfill

\newpage
\spacingset{1.8} 

\section{Introduction}

Statistics on manifolds provides tools for data objects whose natural sample spaces are curved rather than Euclidean \citep{karcher_1977_RiemannianCenterMass,bhattacharya_2012_NonparametricInferenceManifolds,pennec_2020_RiemannianGeometricStatistics}. Correlation-valued observations are a recurring example. In neuroscience, functional connectivity analyses summarize interactions among brain regions by correlation or network matrices \citep{park_2013_StructuralFunctionalBrain,craddock_2015_ConnectomicsNewApproaches}. Financial econometrics uses time-varying correlations to describe changing dependence among assets \citep{engle_2002_DynamicConditionalCorrelation}. Bayesian multivariate and hierarchical models often decompose a covariance matrix into marginal scales and a correlation matrix, so prior modeling for the correlation component becomes a separate statistical problem \citep{barnard_2000_ModelingCovarianceMatrices,lewandowski_2009_GeneratingRandomCorrelation}. These examples all lead to observations in the open elliptope
\[
\Corr_p=\{C\in \Sym^p:C\succ0,\ \diag(C)=\mathbf 1\},
\]
the manifold of full-rank $p\times p$ correlation matrices. Here $\Sym^p$ denotes the space of real symmetric $p\times p$ matrices, $C\succ0$ means positive definite, and $\diag(C)$ is the vector of diagonal entries. The problem addressed in this paper is to build a normalized likelihood model, together with estimation and sampling algorithms, for observations in $\Corr_p$.

A likelihood model is needed when correlation matrices are more than points to be averaged. Clustering rolling financial correlations requires comparable component likelihoods. Scenario generation requires a sampler that returns valid full-rank correlation matrices. A prior distribution for an unknown correlation matrix should be proper and should have interpretable hyperparameters. These tasks cannot be reduced to manifold-valued summary statistics alone. They require a density, a reference measure, a normalizing constant, and a way to compute or approximate the resulting likelihood.

The geometry of correlation matrices is dictated by the removal of marginal scale. If $\Sigma$ is a covariance matrix and $D$ is a positive diagonal matrix, then $\Sigma$ and $D\Sigma D$ have the same correlation matrix. A full-rank correlation matrix is therefore an equivalence class in the symmetric positive definite (SPD) cone $\SPD^p$ modulo positive diagonal congruence. The affine-invariant metric on $\SPD^p$ is invariant under congruence transformations, and the quotient by positive diagonal scaling gives the quotient-affine geometry on $\Corr_p$ \citep{david_2019_RiemannianStructureCorrelation, thanwerdas_2021_GeodesicQuotientAffineMetrics}. We use this quotient structure in every part of the model: the distance, the Riemannian volume, the likelihood, and the sampling algorithms.

Existing approaches answer related but different questions. Euclidean Gaussian models on the off-diagonal entries are simple, but they do not respect the nonlinear geometry of $\Corr_p$. Fisher's transformation is appropriate for a single correlation coefficient \citep{fisher_probable_1921}, but modeling transformed entries separately does not by itself enforce positive definiteness. Unconstrained Cholesky parametrizations of covariance matrices provide useful coordinate systems \citep{pinheiro_1996_UnconstrainedParametrizationsVariancecovariance,pourahmadi_1999_JointMeancovarianceModels}. Hyperspherical constructions offer another route to valid correlation matrices \citep{rebonato_2011_MostGeneralMethodology}. For correlation geometry, the Euclidean-Cholesky metric (ECM) and log-Euclidean-Cholesky metric (LEC) provide computationally attractive baselines \citep{thanwerdas_2022_TheoreticallyComputationallyConvenient}. These models place Gaussian structure in chosen coordinates rather than defining a radial density under the quotient-affine metric. The Lewandowski--Kurowicka--Joe (LKJ) family is a standard determinant-based prior on correlation matrices \citep{lewandowski_2009_GeneratingRandomCorrelation}, but it is not a centered Riemannian Gaussian around an arbitrary correlation template. Riemannian Gaussian distributions on $\SPD^p$ provide the closest precedent \cite{said_2017_RiemannianGaussianDistributions}. A covariance-space Gaussian, however, still treats marginal scale as part of the object, whereas a correlation-space model must remove that scale.

We introduce a quotient-affine (QA) Riemannian Gaussian model for full-rank correlation matrices. The theory establishes propriety, derives exact likelihood and profiled-scale equations, identifies the classical Fisher-transform case, and proves center-dependent normalization in dimension three. We also develop numerical methods for distance and volume evaluation, likelihood fitting, and sampling. The simulations stress these methods across dimension and dispersion. The applications compare QA modeling with coordinate mixtures for rolling financial correlations and with an LKJ prior under informative and misspecified centers.

The paper is organized as follows. Section~\ref{sec:geometry} reviews the quotient-affine geometry needed later, with emphasis on results of David--Gu and Thanwerdas--Pennec. Section~\ref{sec:model} defines the Riemannian Gaussian distribution and explains its invariant and coordinate forms. Section~\ref{sec:likelihood} develops exact likelihood inference, including why profiling the scale is useful. Section~\ref{sec:computation} gives numerical integration, likelihood, and sampling procedures. Section~\ref{sec:experiments} reports four simulation studies and presents two applications. The final section interprets the evidence, states limitations, and outlines technical extensions. We record the failure audit, reproducibility details, and all proofs in the Supplementary Material. The source code, simulation scripts, and notebooks required to reproduce
all numerical experiments and applications presented in this article are available at \url{https://github.com/kisungyou/CorrelationRiemGauss}.

\section{Quotient-affine geometry of full-rank correlation matrices}\label{sec:geometry}

\citet{david_2019_RiemannianStructureCorrelation} established the quotient construction used here, followed by development of its Riemannian operations  \citep{thanwerdas_2021_GeodesicQuotientAffineMetrics,thanwerdas_2021_GeodesicsCurvatureQuotientAffine}. This section summarizes only the definitions and operations needed for the likelihood. These include the affine-invariant distance, quotient distance, horizontal lifts, and volume density in a global chart.

\begin{figure}[H]
\centering
\resizebox{0.84\textwidth}{!}{%
\begin{tikzpicture}[
  x=0.72cm,
  y=0.72cm,
  >=Latex,
  every node/.style={font=\footnotesize},
  orbit/.style={draw=black!70,line width=0.75pt},
  mainorbit/.style={draw=black,line width=1.05pt},
  projection/.style={
    ->,
    draw=black!58,
    line width=0.68pt,
    densely dashed,
    shorten <=4pt,
    shorten >=1pt
  },
  point/.style={circle,fill=black,inner sep=1.45pt},
  quotientpoint/.style={circle,fill=black,inner sep=1.55pt}
]

\shade[
  draw=black!85,
  line width=0.95pt,
  top color=white,
  bottom color=black!38,
  shading angle=8
]
  (-5.05,0.72)
  .. controls (-4.20,2.78) and (-2.55,3.66) .. (-0.55,3.88)
  .. controls (1.35,4.08) and (3.12,3.55) .. (4.18,2.43)
  .. controls (4.53,2.06) and (4.56,1.72) .. (4.22,1.49)
  .. controls (4.03,1.36) and (3.88,1.20) .. (3.77,1.02)
  .. controls (1.35,1.50) and (-1.91,1.55) .. (-5.05,0.72)
  -- cycle;

\draw[orbit]
  (-4.02,1.00)
  .. controls (-3.80,2.2) and (-3.28,2.98) .. (-2.43,3.3);
\draw[mainorbit]
  (-1.95,1.50)
  .. controls (-1.93,1.61) and (-1.87,1.73) .. (-1.82,1.84)
  .. controls (-1.67,2.15) and (-1.43,2.47) .. (-1.25,2.67)
  .. controls (-0.98,3.18) and (-0.82,3.36) .. (-0.68,3.50)
  .. controls (-0.61,3.59) and (-0.55,3.70) .. (-0.52,3.82);
\draw[orbit]
  (0.35,1.48)
  .. controls (0.48,2.28) and (0.99,3.10) .. (1.83,3.63);
\draw[orbit]
  (2.62,1.32)
  .. controls (2.53,1.66) and (2.72,2.54) .. (3.30,3.06);

\node[point] (x) at (-1.82,1.84) {};
\node[point] (g2x) at (-1.25,2.67) {};
\node[point] (g1x) at (-0.68,3.50) {};
\node[anchor=east] at (-1.9,1.84) {$x$};
\node[anchor=east] at (-1.3,2.62) {$g_2\!\cdot\!x$};
\node[anchor=east] at (-0.75,3.45) {$g_1\!\cdot\!x$};

\node[anchor=west] at (4.68,2.43) {Ambient manifold $M$};
\node[anchor=west] at (-0.94,2.94) {$G\!\cdot\!x$};

\draw[black!88,line width=1.10pt]
  (-4.75,-1.05)
  .. controls (-2.35,-1.82) and (2.18,-1.82) .. (4.55,-1.06);

\node[quotientpoint] (qleft) at (-3.20,-1.43) {};
\node[quotientpoint] (qx) at (-1.40,-1.63) {};
\node[quotientpoint] (qright) at (0.50,-1.66) {};
\node[quotientpoint] (qfar) at (2.76,-1.48) {};

\draw[projection] (-4.02,1.00) to[out=-88,in=92] (qleft);
\draw[projection] (-1.95,1.50)
  to[out=-88,in=92]
  node[midway,right=2pt] {$\pi$}
  (qx);
\draw[projection] (0.35,1.48) to[out=-88,in=92] (qright);
\draw[projection] (2.62,1.32) to[out=-88,in=92] (qfar);

\node[anchor=north] (classsymbol) at (qx.south) {$[x]$};
\node[anchor=west] at (4.68,-1.08) {Quotient $M/G$};

\end{tikzpicture}%
}
\caption{Schematic quotient structure. Curves across the gray ambient manifold $M$ represent visible portions of group orbits. The marked points $x$, $g_1\!\cdot\!x$, and $g_2\!\cdot\!x$ lie on the same fiber $G\!\cdot\!x$ and therefore project to the single equivalence class $[x]$ in $M/G$. For the correlation problem, $M=\SPD^p$, $G=\Diagp(p)$, and $M/G\simeq\Corr_p$.}
\label{fig:quotient_schematic}
\end{figure}
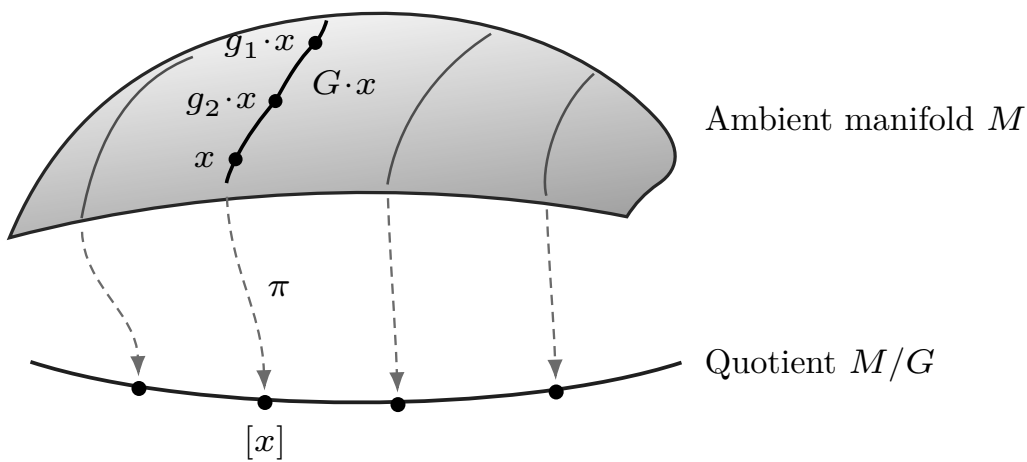

The basic quotient picture is shown schematically in Figure~\ref{fig:quotient_schematic}. The upper space represents an ambient manifold with group orbits, or fibers, consisting of points that should be regarded as equivalent. The lower space represents the quotient, where each fiber is collapsed to a single point. In the correlation problem, the ambient space is the symmetric positive definite cone, the fibers are positive diagonal rescalings of a covariance matrix, and the quotient points are correlation matrices.

\subsection{Affine-invariant geometry on the SPD cone}

The quotient construction starts from the affine-invariant geometry of the SPD cone. We therefore recall its metric, distance, and exponential and logarithm maps before passing to correlation matrices. In this section, we denote by $\exp(\cdot)$ and $\log(\cdot)$ the matrix exponential and logarithm, respectively.

Let $\SPD^p$ be the cone of $p\times p$ symmetric positive definite matrices. The affine-invariant Riemannian metric on $\SPD^p$ is
\[
g^{\AI}_{\Sigma}(U,V)=\tr(\Sigma^{-1}U\Sigma^{-1}V),
\qquad U,V\in T_{\Sigma}\SPD^p\simeq \Sym^p.
\]
The corresponding distance is
\[
d_{\AI}(\Sigma_1,\Sigma_2)=\left\|\log\left(\Sigma_1^{-1/2}\Sigma_2\Sigma_1^{-1/2}\right)\right\|_F.
\]
The exponential and logarithm maps are
\begin{align*}
\Exp^{\AI}_{\Sigma}(U)&=\Sigma^{1/2}\exp\left(\Sigma^{-1/2}U\Sigma^{-1/2}\right)\Sigma^{1/2},\\
\Log^{\AI}_{\Sigma}(A)&=\Sigma^{1/2}\log\left(\Sigma^{-1/2}A\Sigma^{-1/2}\right)\Sigma^{1/2}.
\end{align*}

\subsection{The quotient by positive diagonal scaling}

Correlation matrices discard marginal scales, so the ambient SPD geometry must be quotiented by positive diagonal congruence. The following action and projection make this identification explicit.

Let $\Diagp(p)$ be the group of positive diagonal matrices. It acts on $\SPD^p$ by
\[
D\cdot\Sigma=D\Sigma D.
\]
The projection from covariance matrices to correlation matrices is
\[
\pi(\Sigma)=\Diag(\Sigma)^{-1/2}\Sigma\Diag(\Sigma)^{-1/2},
\]
where $\Diag(\Sigma)$ is the diagonal matrix with the same diagonal as $\Sigma$. The fiber over $C\in\Corr_p$ is
\[
\pi^{-1}(C)=\{DCD:D\in\Diagp(p)\}.
\]
At a correlation representative $C\in\Corr_p$, the differential of $\pi$ is
\begin{equation}\label{eq:dpi_formula}
d\pi_C(U)=U-\frac12\Diag(\diag U)C-\frac12 C\Diag(\diag U),
\qquad U\in\Sym^p.
\end{equation}
The affine-invariant metric is invariant under congruence by any nonsingular matrix, hence in particular under positive diagonal congruence. Following \citet{david_2019_RiemannianStructureCorrelation}, it induces a quotient metric on $\SPD^p/\Diagp(p)$, identified with $\Corr_p$.

The quotient-affine distance between two full-rank correlation matrices is
\begin{equation}\label{eq:qa_distance}
\rho(C_1,C_2)=d_{\QA}(C_1,C_2)=\inf_{D\in\Diagp(p)}d_{\AI}(C_1,DC_2D).
\end{equation}
Equivalently, with $D_{\delta}=\Diag(e^{\delta_1},\ldots,e^{\delta_p})$,
\begin{equation}\label{eq:qa_distance_delta}
\rho(C_1,C_2)^2=\inf_{\delta\in\R^p}\left\|\log\left(C_1^{-1/2}D_{\delta}C_2D_{\delta}C_1^{-1/2}\right)\right\|_F^2.
\end{equation}
Equations \eqref{eq:qa_distance} and \eqref{eq:qa_distance_delta} are standard consequences of the quotient construction. They are included because every later likelihood evaluation, sampler, and approximation uses this distance as a primitive. The objective in \eqref{eq:qa_distance_delta} is smooth in $\delta$. We do not assert that it is globally convex. Nonunique minimizers are the numerical counterpart of nonunique quotient logarithms, leading to algorithms with multistart and stationarity diagnostics.

\subsection{Horizontal lifts, quotient logarithms, and volume density}

Likelihood computation requires tangent operations and a volume element on the quotient rather than on the ambient SPD cone. Horizontal representatives provide both quantities by separating geometric directions from changes in marginal scale.

At $C\in\Corr_p$, a tangent vector is a symmetric matrix $H$ with zero diagonal. A tangent vector to the fiber through $C$ has the form $AC+CA$, where $A$ is diagonal. The horizontal space is the $g^{\AI}$-orthogonal complement of this vertical space. This decomposition connects the ambient SPD geometry to the quotient geometry. It identifies the ambient tangent vector that represents a given direction on the correlation manifold.

\begin{lemma}[Horizontal lift]\label{lem:horizontal_lift}
Let $C\in\Corr_p$ and let $H\in T_C\Corr_p$, so $H=H^T$ and $\diag(H)=0$. Define $B(C)=C^{-1}\Had C$, that is,
\[
B(C)_{ij}=(C^{-1})_{ij}C_{ij}.
\]
Let $a\in\R^p$ solve
\[
\{I+B(C)\}a=-\diag(C^{-1}H).
\]
With $A=\Diag(a)$, define
\[
H^{\mathrm H}=H+AC+CA.
\]
Then $d\pi_C(H^{\mathrm H})=H$, and $H^{\mathrm H}$ is horizontal:
\[
\diag(C^{-1}H^{\mathrm H})=0.
\]
\end{lemma}

The quotient-affine logarithm is defined at points where the minimizing endpoint representative is unique. If $D^*(\bar C,C)$ is the unique minimizer of \eqref{eq:qa_distance} with base point $\bar C$, set
\begin{equation}\label{eq:qa_log_definition}
\Log^{\QA}_{\bar C}(C)=d\pi_{\bar C}\left\{\Log^{\AI}_{\bar C}\bigl(D^*(\bar C,C)\,C\,D^*(\bar C,C)\bigr)\right\}.
\end{equation}
Equivalently, $\Log^{\QA}_{\bar C}(C)$ is the initial velocity of the minimizing quotient geodesic from $\bar C$ to $C$. Nonuniqueness occurs on the cut locus and is a null-set issue for absolutely continuous distributions under the conditions in Proposition \ref{prop:geometric_foundation}.

Let $x\mapsto C(x)$ be a smooth global chart for $\Corr_p$. We use the normalized-Cholesky chart for computation. Let $L(C)$ be the lower-triangular Cholesky factor of $C$ with positive diagonal, and define
\[
\Theta(C)=\Diag(L(C))^{-1}L(C).
\]
Then $\Theta(C)$ is unit lower triangular. The chart is
\begin{equation}\label{eq:cholesky_chart}
x=\vecl\{\Theta(C)-I\}\in\R^d,
\qquad d=p(p-1)/2,
\end{equation}
where $\vecl$ stacks strict lower-triangular entries. Its inverse maps a unit lower-triangular matrix $\Gamma(x)$ to
\begin{equation}\label{eq:cholesky_chart_inverse}
C(x)=\Diag(\Gamma(x)\Gamma(x)^T)^{-1/2}\Gamma(x)\Gamma(x)^T\Diag(\Gamma(x)\Gamma(x)^T)^{-1/2}.
\end{equation}
This chart is a smooth bijection with smooth inverse. Supplementary Section~\ref{sec:supp_chart} verifies the claim.

Let $E_a(x)=\partial C(x)/\partial x_a$, $a=1,\ldots,d$. Let $E_a^{\mathrm H}(x)$ be the horizontal lift from Lemma \ref{lem:horizontal_lift}. The metric tensor in chart coordinates is
\begin{equation}\label{eq:metric_tensor_chart}
G_{ab}(x)=g^{\AI}_{C(x)}\{E_a^{\mathrm H}(x),E_b^{\mathrm H}(x)\}.
\end{equation}
The quotient-affine volume density in the chart is
\begin{equation}\label{eq:volume_density}
J_{\QA}(x)=\sqrt{\det G(x)}.
\end{equation}

\begin{proposition}[Geometric foundation]\label{prop:geometric_foundation}
The following statements hold for the quotient-affine correlation manifold.
\begin{enumerate}[label=(\alph*)]
\item The action of $\Diagp(p)$ on $(\SPD^p,g^{\AI})$ is free, proper, and isometric. Hence $\SPD^p/\Diagp(p)$ is a smooth quotient manifold and $\pi$ is a Riemannian submersion onto $\Corr_p$.
\item The quotient-affine metric is complete. The infimum in \eqref{eq:qa_distance} is attained for every pair $C_1,C_2\in\Corr_p$.
\item For each $\bar C\in\Corr_p$, the cut locus has quotient-affine volume zero. Thus $C\mapsto\rho(C,\bar C)^2$ is smooth outside a null set.
\item There exist constants $A_p,b_p<\infty$ such that, for all $\bar C\in\Corr_p$ and all $r\ge0$,
\begin{equation}\label{eq:volume_growth}
\vol_{\QA}\{B_{\QA}(\bar C,r)\}\le A_p e^{b_p r},
\end{equation}
where $B_{\QA}(\bar C,r)$ is the quotient-affine metric ball of radius $r$ around $\bar C$. Consequently, for every $\sigma>0$ and every integer $k\ge0$,
\[
\int_{\Corr_p}\rho(C,\bar C)^k\exp\left\{-\frac{\rho(C,\bar C)^2}{2\sigma^2}\right\}d\vol_{\QA}(C)<\infty.
\]
\end{enumerate}
\end{proposition}

The proposition justifies treating the quotient-affine distance as a metric quantity rather than a formal infimum over covariance representatives. Its volume bound also supplies the tail control needed for normalized densities, likelihood differentiation, and sampling.

\section{The quotient-affine Riemannian Gaussian distribution}\label{sec:model}

The model combines a radial kernel based on quotient-affine distance with quotient-affine volume. We first establish its invariant and coordinate forms. A maximum-entropy characterization explains the radial choice. Two special cases then clarify the likelihood: Fisher-transform normality when $p=2$ and center-dependent normalization when $p=3$.

\subsection{Model definition and quotient invariance}

The model is centered at a full-rank correlation matrix $\bar C\in\Corr_p$ and has a positive scale parameter $\sigma>0$. Its density is written with respect to quotient-affine Riemannian volume, denoted $d\vol_{\QA}(C)$. The normalizing constant is denoted by $Z(\bar C,\sigma)$ and is part of the likelihood.

\begin{definition}[Quotient-affine Riemannian Gaussian]\label{def:rg}
Let $\bar C\in\Corr_p$ and $\sigma>0$. The quotient-affine Riemannian Gaussian distribution, denoted
\[
C\sim\NR(\bar C,\sigma),
\]
is the probability distribution on $\Corr_p$ with density
\begin{equation}\label{eq:rg_density}
p(C\mid \bar C,\sigma)=\frac{1}{Z(\bar C,\sigma)}\exp\left\{-\frac{\rho(C,\bar C)^2}{2\sigma^2}\right\}
\end{equation}
with respect to $d\vol_{\QA}(C)$, where
\begin{equation}\label{eq:normalizing_constant}
Z(\bar C,\sigma)=\int_{\Corr_p}\exp\left\{-\frac{\rho(C,\bar C)^2}{2\sigma^2}\right\}d\vol_{\QA}(C).
\end{equation}
\end{definition}
Expectations under $\NR(\bar C,\sigma)$ are denoted by $\E_{\bar C,\sigma}$.

The finiteness of $Z(\bar C,\sigma)$ is not an assumption. Proposition \ref{prop:geometric_foundation} proves exponential volume growth and finite radial moments, which imply that \eqref{eq:normalizing_constant} is finite for every center and every positive scale. Thus the distribution is a well-defined probability model before any numerical approximation is introduced.

The definition is also independent of arbitrary covariance representatives. If $[\Sigma]$ denotes the equivalence class of $\Sigma\in\SPD^p$ under positive diagonal congruence, the distance between two classes is obtained by optimizing over endpoint representatives. The following proposition records that this representative choice does not affect the density.

\begin{proposition}[Well-defined quotient density]\label{prop:well_defined}
Let $[\Sigma]$ denote the equivalence class of $\Sigma\in\SPD^p$ under positive diagonal congruence. The function
\[
([\Sigma],[\bar\Sigma])\mapsto \inf_{D\in\Diagp(p)}d_{\AI}(\Sigma,D\bar\Sigma D)
\]
is independent of the representatives $\Sigma$ and $\bar\Sigma$. Consequently, Definition \ref{def:rg} defines a density on $\SPD^p/\Diagp(p)$, equivalently on $\Corr_p$, rather than on a particular covariance representative.
\end{proposition}

Representative invariance removes any dependence on arbitrary marginal variances. The coordinates below are introduced solely for computation and do not redefine the model.

\subsection{Coordinate density for computation}

The invariant density in Definition \ref{def:rg} is the conceptual model. Numerical likelihood evaluation and sampling require a coordinate representation. We use the normalized-Cholesky chart introduced in \eqref{eq:cholesky_chart}--\eqref{eq:cholesky_chart_inverse}. This chart maps a vector $x\in\R^d$ to a correlation matrix $C(x)$, where $d=p(p-1)/2$. The Riemannian volume element is not Lebesgue measure in $x$, but contributes the factor $J_{\QA}(x)=\sqrt{\det G(x)}$ where $G(x)$ is the quotient-affine metric tensor from \eqref{eq:metric_tensor_chart}. The next proposition converts the invariant density into the coordinate form used by the algorithms.

\begin{proposition}[Coordinate density]\label{prop:coordinate_density}
Let $x\mapsto C(x)$ be the normalized-Cholesky global chart. For the random vector $X=x(C)$, the density of $X$ with respect to Lebesgue measure on $\R^d$ is
\begin{equation}\label{eq:coordinate_density}
p_X(x\mid \bar C,\sigma)=\frac{1}{Z(\bar C,\sigma)}\exp\left\{-\frac{\rho(C(x),\bar C)^2}{2\sigma^2}\right\}J_{\QA}(x),
\end{equation}
where $J_{\QA}(x)=\sqrt{\det G(x)}$ is given in \eqref{eq:volume_density}. Therefore
\begin{equation}\label{eq:chart_integral_Z}
Z(\bar C,\sigma)=\int_{\R^d}\exp\left\{-\frac{\rho(C(x),\bar C)^2}{2\sigma^2}\right\}J_{\QA}(x)\,dx.
\end{equation}
\end{proposition}

This coordinate formula also clarifies the distinction between the proposed model and a Cholesky-coordinate Gaussian. A Cholesky-coordinate Gaussian would place a normal density directly on $x$. In contrast, the density in \eqref{eq:coordinate_density} is the quotient-affine radial density expressed in the $x$ chart. The Jacobian factor $J_{\QA}$ is therefore essential. In the algorithms below, every numerical integral for $Z$ is an integral of \eqref{eq:coordinate_density}. Omitting $J_{\QA}$ would change the reference measure and hence change the statistical model.

\subsection{Maximum-entropy characterization}

The radial form is not chosen only by analogy. Fixing a center $\bar C$ and  the expected squared quotient-affine radius, the density in Definition \ref{def:rg} is the maximum-entropy density with respect to quotient-affine volume. This gives a variational reason to pay the computational cost of the normalizing constant: the model is the Gibbs distribution associated with squared Riemannian distance.

\begin{proposition}[Maximum-entropy characterization]\label{prop:max_entropy}
Fix $\bar C\in\Corr_p$ and suppose $\sigma>0$ is chosen so that $m=\E_{\bar C,\sigma}\{\rho(C,\bar C)^2\}$ is finite. Among all densities $q$ with respect to $d\vol_{\QA}$ satisfying
\[
\int q(C)d\vol_{\QA}(C)=1,
\qquad
\int \rho(C,\bar C)^2q(C)d\vol_{\QA}(C)=m,
\]
the density in \eqref{eq:rg_density} maximizes entropy $-\int q\log q\,d\vol_{\QA}$.
\end{proposition}

Thus the radial family is the least informative density, relative to quotient-affine volume, at a fixed expected squared radius. We contrast this choice with possible pushforward and wrapped constructions.

\subsection{Exact two-dimensional case}

The case $p=2$ provides an analytic check on the geometry, likelihood, and sampler. A correlation matrix of size $2\times2$  is determined by a single correlation $r\in(-1,1)$ in its off-diagonal, and the quotient-affine coordinate turns out to be Fisher's transformation.

\begin{proposition}[The $2\times2$ case]\label{prop:p2}
For
\[
C(r)=\begin{pmatrix}1&r\\ r&1\end{pmatrix},\qquad r\in(-1,1),
\]
let $z=\operatorname{atanh}(r)$. Then
\begin{equation}\label{eq:p2_distance}
\rho\{C(r_1),C(r_2)\}=\sqrt{2}\,|z_1-z_2|.
\end{equation}
The quotient-affine line element and volume element are
\[
ds^2=2\,dz^2,
\qquad
 d\vol_{\QA}=\sqrt2\,dz.
\]
Consequently,
\[
Z\{C(r_0),\sigma\}=\sqrt{2\pi}\,\sigma,
\]
which is independent of the center. Equivalently, under $C(r)\sim\NR(C(r_0),\sigma)$,
\[
z=\operatorname{atanh}(r)\sim N(z_0,\sigma^2/2),\qquad z_0=\operatorname{atanh}(r_0).
\]
\end{proposition}

This proposition is the simplest check that the construction has the right classical limit. In the scalar-correlation case, the quotient-affine radius is exactly a constant multiple of distance in Fisher's $z$ coordinate, and the Riemannian volume contributes only a constant factor. The likelihood is therefore the ordinary normal likelihood in $z$.

\begin{corollary}[Fisher-transform MLE]\label{cor:p2_mle}
For $p=2$, if $C_i=C(r_i)$ and $z_i=\operatorname{atanh}(r_i)$, then the exact MLE under $\NR(C(r_0),\sigma)$ is
\begin{equation}\label{eq:p2_mle}
\widehat z_0=\frac1n\sum_{i=1}^n z_i,
\qquad
\widehat\sigma^2=\frac{2}{n}\sum_{i=1}^n(z_i-\widehat z_0)^2,
\qquad
\widehat r_0=\tanh(\widehat z_0).
\end{equation}
The exact MLE and the Fr\'echet estimator coincide in this flat one-dimensional case.
\end{corollary}

Center independence at $p=2$ eliminates the likelihood drift term. The next result shows that this simplification fails at $p=3$.

\subsection{Center dependence of the normalizing constant}\label{sec:center_dependence}

The SPD cone with the affine-invariant metric is homogeneous, which is why the SPD Riemannian Gaussian of \citet{said_2017_RiemannianGaussianDistributions} has a center-free normalizing constant. The quotient-affine correlation manifold is different. The congruences that normalize the positive diagonal group are monomial transformations, which act by permutations and sign change after projection. This group fixes the identity correlation matrix and is not transitive on $\Corr_p$.

Nontransitivity alone does not prove that $Z$ depends on the center. The next proposition combines an explicit scalar-curvature calculation with the small-dispersion expansion in Proposition \ref{prop:laplace}. That expansion is stated in Section~\ref{sec:computation}, and its proof is independent of this section. The calculation explains why exact MLE can differ from Fr\'echet estimation beyond the $2\times2$ validation case.

\begin{proposition}[Center-dependence for $p=3$]\label{prop:center_dependence}
For $p=3$, the scalar curvature of the quotient-affine metric is not constant. In the normalized-Cholesky chart, along the curve
\[
\Gamma_a=\begin{pmatrix}1&0&0\\ a&1&0\\0&0&1\end{pmatrix},
\qquad
C_a=\Diag(\Gamma_a\Gamma_a^T)^{-1/2}\Gamma_a\Gamma_a^T\Diag(\Gamma_a\Gamma_a^T)^{-1/2},
\]
one has
\begin{equation}\label{eq:scalar_curve}
\Scal(C_a)=-\frac{3}{4(1+a^2)}.
\end{equation}
Equivalently, if $r=a/\sqrt{1+a^2}$ is the nonzero block correlation in $C_a$, then $\Scal(C_a)=-\tfrac34(1-r^2)$. In particular, $\Scal(C_0)=-3/4$ and $\Scal(C_1)=-3/8$. Therefore $Z(\bar C,\sigma)$ is center-dependent for all sufficiently small $\sigma>0$ in dimension $p=3$.
\end{proposition}

This calculation establishes center dependence for $p=3$ at small dispersion. Its behavior in larger dimensions remains open, and the experiments examine $Z$ beyond the analytic case.

\begin{corollary}[Population Fr\'echet drift]\label{cor:frechet_drift}
Let $C\sim\NR(C_0,\sigma_0)$ and suppose quotient logarithms are almost surely unique. If
\[
\nabla_{C_0}\log Z(C_0,\sigma_0)\ne0,
\]
then $C_0$ is not a stationary point of the population Fr\'echet objective
\[
F(Q)=\E_{C_0,\sigma_0}\{\rho(C,Q)^2\}.
\]
Consequently, when $F$ has a unique minimizer, the population Fr\'echet mean differs from the model center $C_0$. 
\end{corollary}

\section{Likelihood-based inference}\label{sec:likelihood}

The previous section defined the model and showed why the normalizing constant cannot be ignored in general. We now turn that observation into an estimation procedure. The likelihood is called exact because it keeps the full normalizing constant $Z(\bar C,\sigma)$. Later numerical schemes approximate this likelihood, not changing the statistical model.

\subsection{Exact likelihood and score equations}

Center-dependent normalization changes both center and scale estimation. We therefore write the full likelihood before deriving its score equations.

For independent observations $C_1,\ldots,C_n\in\Corr_p$, define
\[
S_n(\bar C)=\sum_{i=1}^n \rho(C_i,\bar C)^2.
\]
The exact log-likelihood is
\begin{equation}\label{eq:loglik}
\ell(\bar C,\sigma)=-\frac{S_n(\bar C)}{2\sigma^2}-n\log Z(\bar C,\sigma).
\end{equation}
The exact MLE is any maximizer of \eqref{eq:loglik}. Since $Z(\bar C,\sigma)$ may depend on $\bar C$, the MLE is not generally the minimizer of $S_n(\bar C)$. The score equations below make this difference explicit. They also explain why the same numerical integration used for $Z$ is needed for likelihood-based optimization.

\begin{theorem}[Exact score equations]\label{thm:score}
Fix $(\bar C,\sigma)$ and suppose the quotient logarithms $\Log^{\QA}_{\bar C}(C_i)$ are uniquely defined. Then the center score satisfies
\begin{equation}\label{eq:center_score}
\nabla_{\bar C}\ell(\bar C,\sigma)=\frac{1}{\sigma^2}\left[\sum_{i=1}^n \Log^{\QA}_{\bar C}(C_i)-n\E_{\bar C,\sigma}\{\Log^{\QA}_{\bar C}(C)\}\right].
\end{equation}
Hence any interior exact MLE satisfies
\begin{equation}\label{eq:center_score_eq}
\frac1n\sum_{i=1}^n \Log^{\QA}_{\widehat C}(C_i)=\E_{\widehat C,\widehat\sigma}\{\Log^{\QA}_{\widehat C}(C)\}.
\end{equation}
The scale score is
\begin{equation}\label{eq:scale_score}
\frac{\partial}{\partial\sigma}\ell(\bar C,\sigma)=\frac{S_n(\bar C)}{\sigma^3}-n\frac{\E_{\bar C,\sigma}\{\rho(C,\bar C)^2\}}{\sigma^3},
\end{equation}
so an interior MLE also satisfies
\begin{equation}\label{eq:scale_score_eq}
\frac1nS_n(\widehat C)=\E_{\widehat C,\widehat\sigma}\{\rho(C,\widehat C)^2\}.
\end{equation}
\end{theorem}

The \Frechet estimating equation is obtained by replacing the right-hand side of \eqref{eq:center_score_eq} by zero. Thus Fr\'echet estimation ignores the normalizing-constant drift term
\[
\E_{\bar C,\sigma}\{\Log^{\QA}_{\bar C}(C)\}=\sigma^2\nabla_{\bar C}\log Z(\bar C,\sigma).
\]

\subsection{Profile likelihood for the scale parameter}

The exact likelihood involves a center on a curved space and a positive scale. Joint optimization is possible, but each trial scale requires another normalizer evaluation or interpolation. For a fixed center, the scale equation becomes one-dimensional and monotone. Profiling preserves the exact likelihood while reducing the outer search to the center and replacing joint scale optimization by one-dimensional root finding.

\begin{proposition}[Profile scale equation]\label{prop:profile_scale}
For fixed $\bar C$, let
\[
m_2(\bar C,\sigma)=\E_{\bar C,\sigma}\{\rho(C,\bar C)^2\}.
\]
Then an exact scale MLE for fixed $\bar C$ solves
\[
m_2(\bar C,\sigma)=\frac{S_n(\bar C)}{n}.
\]
Moreover,
\begin{equation}\label{eq:m2_monotone}
\frac{\partial}{\partial\sigma}m_2(\bar C,\sigma)=\frac{\Var_{\bar C,\sigma}\{\rho(C,\bar C)^2\}}{\sigma^3}\ge0.
\end{equation}
The inequality is strict unless the squared radius is degenerate.
\end{proposition}

This profiled form is not a different estimator, but a computational representation of the same exact likelihood with the scale chosen to satisfy the likelihood equation for each candidate center.

The monotonicity in Proposition \ref{prop:profile_scale} supports one-dimensional root finding in $\sigma$ for each fixed center. More precisely, $m_2(\bar C,\sigma)$ is continuous and nondecreasing, with $m_2(\bar C,\sigma)\downarrow0$ as $\sigma\downarrow0$. Let $m_2^{\sup}(\bar C)=\lim_{\sigma\uparrow\infty}m_2(\bar C,\sigma)\in(0,\infty]$. The profile equation is solved only when $S_n(\bar C)/n<m_2^{\sup}(\bar C)$; in computation this condition is checked on the scale grid used to approximate $m_2$. Whether $m_2^{\sup}(\bar C)=\infty$ for all $p\ge3$ is not needed for the algorithms and is left as a geometric question.

\subsection{Identifiability and consistency}

After profiling, the remaining population question is whether distinct parameters determine distinct densities. If two parameter values gave the same density, no amount of data could distinguish them.

\begin{proposition}[Identifiability]\label{prop:identifiability}
If
\[
p(\cdot\mid C_1,\sigma_1)=p(\cdot\mid C_2,
\sigma_2)
\]
as densities with respect to $d\vol_{\QA}$, then $C_1=C_2$ and $\sigma_1=\sigma_2$.
\end{proposition}

Identifiability helps to fix the population target. Consistency additionally requires uniform convergence of the sample log likelihood over the searched parameter region. On compact center sets and bounded scale intervals, continuity and Proposition~\ref{prop:geometric_foundation} provide an integrable envelope. The following result concerns exact likelihood on compact parameter sets, not global optimization over the noncompact manifold.

\begin{theorem}[Consistency on compact parameter sets]\label{thm:mle_consistency}
Let $C_1,C_2,\ldots$ be independent and identically distributed from $\NR(C_0,\sigma_0)$. Let $\Theta=K\times[\sigma_{\min},\sigma_{\max}]$, where $K\subset\Corr_p$ is compact, $0<\sigma_{\min}<\sigma_0<\sigma_{\max}<\infty$, and $(C_0,\sigma_0)\in\Theta$. If the exact log-likelihood is maximized over $\Theta$, then every sequence of exact MLEs converges in probability to $(C_0,\sigma_0)$.
\end{theorem}

We state the theorem on compact parameter sets because the paper concerns computational likelihood evaluation rather than global optimization theory. In applications, this corresponds to a numerically inspected center region and scales bounded away from zero and infinity.

We do not claim asymptotic normality here. Standard M-estimation theory would additionally require an interior true parameter, a nonsingular information matrix, and locally uniform control of the Hessian. Proposition~\ref{prop:geometric_foundation} supplies the needed tail envelopes on compact neighborhoods. Nonsingularity and second-order regularity would still require separate verification.

\section{Computation}\label{sec:computation}

Exact likelihood evaluation and sampling require numerical access to the normalizing constant and the quotient geometry. This section develops a local approximation, integration strategies, core algorithms, complexity bounds, and sampler guarantees.

\subsection{Local Laplace approximation}

The leading approximation to $Z$ is obtained from normal coordinates at $\bar C$. The result is standard, but it is stated here because it explains both the Fr\'echet approximation and the center-dependence mechanism.

\begin{proposition}[Small-dispersion expansion]\label{prop:laplace}
Fix $\bar C\in\Corr_p$ and choose $r_0<\operatorname{inj}(\bar C)$, where $\operatorname{inj}(\bar C)$ is the injectivity radius at $\bar C$. Then, as $\sigma\downarrow0$,
\begin{equation}\label{eq:laplace_Z}
Z(\bar C,\sigma)=(2\pi\sigma^2)^{d/2}\left\{1-\frac{\Scal(\bar C)}{6}\sigma^2+O_{\bar C}(\sigma^4)\right\}.
\end{equation}
The remainder is uniform for $\bar C$ in compact subsets with injectivity radius bounded below.
\end{proposition}

The leading term gives the Fr\'echet approximation
\begin{equation}\label{eq:frechet_estimator}
\widehat C_F=\argmin_{C\in\Corr_p}\sum_{i=1}^n\rho(C_i,C)^2,
\qquad
\widehat\sigma_F^2=\frac1{nd}\sum_{i=1}^n\rho(C_i,\widehat C_F)^2.
\end{equation}
A curvature-corrected approximation replaces $\log Z$ by the logarithm of \eqref{eq:laplace_Z}. This is used as an approximation, not as an exact likelihood.

\subsection{Numerical integration and proposal construction}\label{sec:integration_regimes}

The chart integral in \eqref{eq:chart_integral_Z} can be evaluated at several levels of numerical cost. For $d\le6$ we recommend deterministic adaptive cubature or sparse-grid quadrature as the default benchmark. For moderate dimensions, approximately $6<d\le15$, randomized quasi-Monte Carlo (RQMC) rules provide smoother likelihood surfaces than ordinary Monte Carlo when common random numbers are fixed across nearby parameter values. For larger dimensions, we use heavy-tailed importance sampling. The proposal is a multivariate $t_\nu$ distribution centered at $x(\bar C)$, with scale proportional to $G(x(\bar C))^{-1}$. Pilot effective sample size (ESS) calculations determine any scalar inflation. The same proposal estimates $Z$, $m_2$, and the expectations in the score equations. These choices follow standard Monte Carlo, sampling-importance-resampling, and RQMC practice \citep{robert_1999_MonteCarloStatistical, rubin_1988_UsingSIRAlgorithm, smith_1992_BayesianStatisticsTears}. In all regimes we report the integration rule, number of nodes or draws, Monte Carlo standard error when applicable, ESS, and sensitivity to proposal tail thickness.

\subsection{Likelihood and sampling algorithms}\label{sec:algorithms}

The algorithms are organized around three primitives. Algorithm~\ref{alg:distance} revisits the quotient-affine distance and, when the minimizing representative is isolated, the quotient logarithm, originally proposed by \citet{david_2019_RiemannianStructureCorrelation}. Algorithm~\ref{alg:volume_density} computes the Riemannian volume density in the computational chart. Algorithm~\ref{alg:exact_mle} combines these primitives with an integration rule for exact likelihood fitting. For completeness, we introduce the two samplers that use the same chart density, making the likelihood and sampling code paths to share the same geometric quantities.

\begin{algorithm}[ht]
\caption{Quotient-affine distance and logarithm}
\label{alg:distance}
\begin{algorithmic}[1]
\Require $C_1,C_2\in\Corr_p$; tolerance $\varepsilon_{\delta}$; multistart set $\mathcal D_0\subset\R^p$.
\Ensure Distance $\rho(C_1,C_2)$ and, if unique, $\Log^{\QA}_{C_1}(C_2)$.
\For{$\delta^{(0)}\in\mathcal D_0$}
    \State Minimize $f(\delta)=\|\log(C_1^{-1/2}D_{\delta}C_2D_{\delta}C_1^{-1/2})\|_F^2$ using gradient-based optimization.
    \State Stop when $\|\nabla f(\delta)\|\le\varepsilon_{\delta}(1+f(\delta))$ and relative objective change is below $\varepsilon_{\delta}$.
\EndFor
\State Let $\delta^*$ be the best local minimizer found; set $D^*=D_{\delta^*}$.
\State $\rho(C_1,C_2)\leftarrow f(\delta^*)^{1/2}$.
\State If the minimizer is isolated, set $\Log^{\QA}_{C_1}(C_2)=d\pi_{C_1}\{\Log^{\AI}_{C_1}(D^*C_2D^*)\}$.
\State Report the multistart gap and stationarity residual.
\end{algorithmic}
\end{algorithm}

The derivative of $f$ in Algorithm \ref{alg:distance} can be evaluated using
\[
\frac{\partial f}{\partial\delta_j}=2\tr\left[\log A(\delta)A(\delta)^{-1}\frac{\partial A(\delta)}{\partial\delta_j}\right],
\]
where $A(\delta)=C_1^{-1/2}D_{\delta}C_2D_{\delta}C_1^{-1/2}$ and
\[
\frac{\partial A(\delta)}{\partial\delta_j}=C_1^{-1/2}\{E_jD_{\delta}C_2D_{\delta}+D_{\delta}C_2D_{\delta}E_j\}C_1^{-1/2},
\]
with $E_j$ the diagonal matrix having a one in entry $j$. When $\delta^*$ is unique, Danskin's envelope theorem implies that derivatives of $\rho^2$ with respect to chart variables or the center can be computed by differentiating $f$ at $\delta^*$ without differentiating through the argmin.

\begin{algorithm}[H]
\caption{Quotient-affine volume density in a chart}
\label{alg:volume_density}
\begin{algorithmic}[1]
\Require Chart coordinate $x\in\R^d$; chart map $C(x)$.
\Ensure $J_{\QA}(x)$.
\State Compute $C=C(x)$, $C^{-1}$, and coordinate tangents $E_a=\partial C(x)/\partial x_a$, $a=1,\ldots,d$.
\State Form $B(C)=C^{-1}\Had C$ and factor $I+B(C)$.
\For{$a=1,\ldots,d$}
    \State Solve $(I+B(C))u_a=-\diag(C^{-1}E_a)$.
    \State Set $A_a=\Diag(u_a)$ and $E_a^{\mathrm H}=E_a+A_aC+CA_a$.
\EndFor
\State Form $G_{ab}=\tr(C^{-1}E_a^{\mathrm H}C^{-1}E_b^{\mathrm H})$.
\State Return $J_{\QA}(x)=\sqrt{\det G}$.
\end{algorithmic}
\end{algorithm}

\begin{algorithm}[H]
\caption{Exact MLE by profiled likelihood}
\label{alg:exact_mle}
\begin{algorithmic}[1]
\Require Data $C_1,\ldots,C_n$; chart $\theta\mapsto C(\theta)$; integration rule for $Z$ and $m_2$; tolerances $\varepsilon_{\ell},\varepsilon_{\sigma}$.
\Ensure Approximate exact MLE $(\widehat C_{\mathrm{MLE}},\widehat\sigma_{\mathrm{MLE}})$.
\State Initialize $\theta^{(0)}$ by the Fr\'echet estimator or a coordinate mean; initialize $\sigma^{(0)}$ by \eqref{eq:frechet_estimator}.
\For{outer iterations $t=0,1,2,\ldots$}
    \State Compute $S_n(C(\theta^{(t)}))$ using Algorithm \ref{alg:distance}, warm-starting inner distance solves.
    \State Find $\sigma(\theta^{(t)})$ by solving $m_2(C(\theta^{(t)}),\sigma)=S_n(C(\theta^{(t)}))/n$ to tolerance $\varepsilon_{\sigma}$.
    \State Evaluate the profiled log-likelihood $\ell_{\mathrm{prof}}(\theta^{(t)})$ using \eqref{eq:loglik} and \eqref{eq:chart_integral_Z}.
    \State Update $\theta$ by Broyden--Fletcher--Goldfarb--Shanno, trust-region, or derivative-free optimization with common random numbers when Monte Carlo integration is used.
    \State Terminate when relative log-likelihood change and step size are below $\varepsilon_{\ell}$.
\EndFor
\State Return $\widehat C_{\mathrm{MLE}}=C(\widehat\theta)$ and $\widehat\sigma_{\mathrm{MLE}}=\sigma(\widehat\theta)$ with numerical-integration diagnostics.
\end{algorithmic}
\end{algorithm}

\begin{algorithm}[H]
\caption{Metropolis--Hastings sampler for $\NR(\bar C,\sigma)$}
\label{alg:mh_sampler}
\begin{algorithmic}[1]
\Require Center $\bar C$, scale $\sigma$, starting chart point $x^{(0)}$, proposal covariance $\Sigma_q$, proposal scale $s$, number of iterations $M$.
\Ensure Markov chain $C^{(m)}=C(x^{(m)})$.
\For{$m=1,\ldots,M$}
    \State Propose $x^*=x^{(m-1)}+\varepsilon$, $\varepsilon\sim N(0,s^2\Sigma_q)$.
    \State Compute $\log\pi(x^*)=-\rho(C(x^*),\bar C)^2/(2\sigma^2)+\log J_{\QA}(x^*)$.
    \State Compute $\log\pi(x^{(m-1)})$ analogously.
    \State Accept $x^*$ with probability $\min\{1,\exp(\log\pi(x^*)-\log\pi(x^{(m-1)}))\}$; otherwise keep $x^{(m)}=x^{(m-1)}$.
\EndFor
\end{algorithmic}
\end{algorithm}

\begin{algorithm}[H]
\caption{Sampling-importance-resampling (SIR) sampler and normalizing-constant estimator}
\label{alg:sir_sampler}
\begin{algorithmic}[1]
\Require Center $\bar C$, scale $\sigma$, proposal density $q$ on $\R^d$, proposal sample size $M$, resample size $B$.
\Ensure Approximate independent samples and $\widehat Z$.
\For{$m=1,\ldots,M$}
    \State Draw $x_m\sim q$ and compute
    \[
    w_m=\frac{\exp\{-\rho(C(x_m),\bar C)^2/(2\sigma^2)\}J_{\QA}(x_m)}{q(x_m)}.
    \]
\EndFor
\State Estimate $\widehat Z=M^{-1}\sum_{m=1}^M w_m$ and $\ess=(\sum_m w_m)^2/\sum_m w_m^2$.
\State Normalize $\widetilde w_m=w_m/\sum_{\ell=1}^M w_{\ell}$.
\State Draw $I_1,\ldots,I_B$ from the categorical distribution with probabilities $\widetilde w_1,\ldots,\widetilde w_M$.
\State Return $C(x_{I_1}),\ldots,C(x_{I_B})$, $\widehat Z$, Monte Carlo standard errors, and ESS.
\end{algorithmic}
\end{algorithm}

\subsection{Computational complexity and scalability}\label{sec:complexity}

The computational bottleneck is nested. One evaluation of $\rho(C_1,C_2)$ requires $K_{\delta}$ iterations of the inner optimization in Algorithm \ref{alg:distance} since each iteration requires at least one matrix logarithm or eigendecomposition, costing $O(p^3)$. Thus one distance evaluation costs approximately $O(K_{\delta}p^3)$, plus lower-order linear algebra.

One evaluation of $J_{\QA}(x)$ via Algorithm \ref{alg:volume_density} requires a factorization of $I+B(C)$ at cost $O(p^3)$, $d$ triangular solves, construction of $d$ horizontal lifts, $d^2$ metric inner products, and a determinant of the $d\times d$ metric matrix. Since $d=p(p-1)/2=O(p^2)$, the natural implementation costs $O(p^6)$ per volume-density evaluation after precomputing $C^{-1}E_a^{\mathrm H}$. This $O(p^6)$ term is the main reason exact likelihood is aimed at small to moderate dimensions.

For an outer likelihood optimizer with $T$ iterations, $n$ observations, and $M$ integration points for $Z$, the rough cost is
\begin{equation}\label{eq:complexity}
O\left(T\left[nK_{\delta}p^3+M\{K_{\delta}p^3+p^6\}\right]\right),
\end{equation}
not including multistart factors. Warm starts for $\delta^*$, common random numbers for Monte Carlo likelihood evaluations, and reused factorizations reduce constants but do not change the scaling in \eqref{eq:complexity}. We report wall-clock time, effective sample size, stochastic integration error, and sampler diagnostics in the simulations to appear. Supplementary Section~\ref{sec:supp_failures} gives exception-level completion rates. 

\subsection{Validity and convergence of the sampling algorithms}\label{sec:sampler_correctness}

The sampling algorithms use the chart density in \eqref{eq:coordinate_density}, but their inferential use also requires invariance and convergence guarantees. The next theorem verifies these properties for the Markov chain and importance-resampling procedures.

\begin{theorem}[Sampling correctness]\label{thm:sampler_correctness}
For fixed $(\bar C,\sigma)$, the Metropolis--Hastings chain in Algorithm \ref{alg:mh_sampler} has the chart density in \eqref{eq:coordinate_density} as a stationary density. If the Gaussian random-walk proposal has full support, the chain is $\pi$-irreducible and aperiodic; hence ergodic averages converge for integrable test functions. For Algorithm \ref{alg:sir_sampler}, if $q(x)>0$ whenever the target density is positive, then $\E_q[w(X)]=Z(\bar C,\sigma)$ and self-normalized importance estimates converge almost surely for integrable test functions. For fixed $B$, the resampled empirical distribution converges weakly to the target distribution as $M\to\infty$.
\end{theorem}

The practically binding condition for SIR is not finite first moment, which follows from $Z<\infty$, but adequate tail coverage and a finite or well-controlled second moment of the weights. We monitor this through ESS and sensitivity to proposal tail thickness.

\section{Experiments}\label{sec:experiments}

Computational evidence complements asymptotic complexity when assessing feasibility. Four simulations evaluate implementation accuracy, normalizer computation, normalized-likelihood estimation, and fixed-parameter sampling. Two applications then examine rolling financial correlations and prior information for a latent correlation matrix.

\subsection{Simulation 1: exact two-by-two validation}\label{sec:sim1}

The analytic $p=2$ model provides the only exact benchmark for checking both the likelihood and sampler implementations. We generated 1,000 data sets of size $n=100$ from $z_i\sim N(z_0,\sigma^2/2)$, with $z_0=0.55$ and $\sigma=0.7$, and set $r_i=\tanh(z_i)$. The closed-form estimator in Corollary~\ref{cor:p2_mle} was applied to each data set. Separately, 5,000 exact draws per replication assessed sampler moments, and all 1,000 replications completed.

\begin{table}[H]
\centering
\caption{Two-dimensional likelihood recovery over 1,000 replications with $n=100$. Monte Carlo standard errors (MCSEs) quantify uncertainty in the reported bias across replications. RMSE denotes root mean squared error.}
\label{tab:sim1_recovery}
\begin{tabular}{ccccc}
\toprule
Parameter & Truth & Mean estimate & Bias (MCSE) & RMSE \\
\midrule
$z_0$ & 0.5500 & 0.5524 & \phantom{$-$}0.00243 (0.00158) & 0.0499 \\
$\sigma$ & 0.7000 & 0.6953 & $-0.00467$ (0.00156) & 0.0494 \\
\bottomrule
\end{tabular}
\end{table}

The estimated center and scale had biases 0.00243 and $-0.00467$, with respective RMSEs of 0.0499 and 0.0494. Across exact-sampler replications, the mean was 0.55003 and the variance was 0.24494, compared with targets 0.55 and 0.245. Together, these results validate the closed-form likelihood and sampler implementations, while the reported RMSEs quantify finite-sample variation.

\subsection{Simulation 2: normalizing constant computation}\label{sec:sim2}

For $p>2$, normalizer computation relies on numerical approximation, making accuracy and proposal efficiency central to practical normalized likelihood. We compared the leading and curvature-corrected Laplace expressions with importance sampling and randomized quasi-Monte Carlo (RQMC). The grid included $p\in\{3,4,5\}$, identity, equicorrelation, block-correlation, and near-boundary centers, and $\sigma\in\{0.1,0.25,0.5,1\}$. Each stochastic cell had 20 replications using 32,768 metric-matched $t_5$ proposals and four distance restarts. Each deterministic approximation was evaluated once.

\begin{figure}[ht]
\centering
\includegraphics[width=0.73\textwidth]{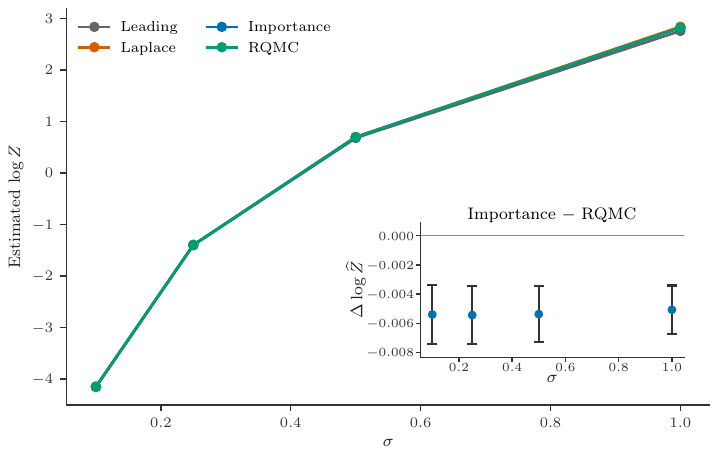}
\caption{Normalizer estimates for the $p=3$ block center. Stochastic points are means across 20 completed randomized replications, while deterministic curves are single evaluations. Because standard errors are smaller than the main-panel markers, the inset magnifies paired importance-minus-RQMC differences. Inset bars show one Monte Carlo standard error of each paired mean difference.}
\label{fig:sim2_logz}
\end{figure}

Importance sampling and RQMC agreed in every fully completed $p=3$ and $p=4$ cell. Their cellwise mean differences in $\log\widehat Z$ ranged from $-0.00632$ to $-0.00428$ for $p=3$ and from 0.00037 to 0.00467 for $p=4$. Beyond the analytic $p=2$ case, this agreement provides an internal check on stochastic normalizer computation. Figure~\ref{fig:sim2_logz} illustrates the comparison for the $p=3$ block center.

The leading expression's mean discrepancy from RQMC changed from $-0.0239$ at $p=3$ to $-0.2667$ at $p=5$, increasing in magnitude with dimension. The latter value is based on completed runs. At the identity, the curvature correction reduced the absolute discrepancy from 0.00126 to 0.00001 at $\sigma=0.1$, while the leading expression was closer at $\sigma=1$. These results locate the correction's clearest accuracy gain at low dispersion. The available correction covers the $p=3$ identity and block families. Extending center-dependent approximations to larger dispersion and higher dimensions is a focused computational direction. Mean stochastic evaluation times ranged from 201 to 254 seconds. The effective-sample-size fraction was about 0.33 at $p=3$ and below 0.08 at $p=5$, quantifying the effect of dimension on proposal efficiency.

\subsection{Simulation 3: normalized likelihood versus Fr\'echet estimation}\label{sec:sim3}

This simulation quantifies the statistical recovery and computational cost associated with normalizer evaluation. Data were generated by SIR, so the experiment evaluates recovery under the implemented target approximation. Simulation~\ref{sec:sim4} provides a complementary cross-sampler assessment. The main grid used $p\in\{3,4\}$, $n\in\{25,50,100,250\}$, $\sigma_0\in\{0.1,0.25,0.5,1\}$, three centers, and 100 planned replications per cell. It compared the Fr\'echet fit with the implemented Laplace-likelihood fit. A smaller $p=3$ grid used three sample sizes, three dispersion levels, two centers, and 30 replications per cell while adding the importance-evaluated likelihood.

Mean center errors were 0.11283 for the Fr\'echet fit and 0.11228 for the Laplace fit. Their paired difference was $-0.000550$, and 66.8\% of paired errors were numerically identical. All but 24 planned fits completed, with the remaining cases confined to $p=4$ and $\sigma_0=1$. The available curvature correction applies to the $p=3$ block family. Other fits use the center-independent leading normalizer. Accordingly, the aggregate results primarily compare leading-normalizer likelihood with Fr\'echet estimation, while the supported block cells isolate the curvature correction. Table~\ref{tab:sim3_importance} reports recovery and fitting times.

\begin{table}[ht]
\centering
\caption{Importance-normalizer subset for $p=3$ over 540 completed replications. Parentheses contain Monte Carlo standard errors of the reported means. Scale error is $|\widehat\sigma-\sigma_0|$.}
\label{tab:sim3_importance}
\begin{tabular}{cccc}
\toprule
Method & Center error & Absolute scale error & Time per fit (s) \\
\midrule
Fr\'echet & 0.06675 (0.00240) & 0.01468 (0.00068) & 26.1 \\
Laplace likelihood & 0.06675 (0.00240) & 0.01468 (0.00068) & 20.8 \\
Importance likelihood & 0.06684 (0.00241) & 0.01506 (0.00068) & 4463.7 \\
\bottomrule
\end{tabular}
\end{table}

On this subset, the three estimators achieved nearly identical center and scale recovery. The importance likelihood supplies a direct-normalizer reference, with an average fit time of 4,463.7 seconds compared with 26.1 seconds for the Fr\'echet fit.

\subsection{Simulation 4: sampler performance}\label{sec:sim4}

Finite-run efficiency and preservation of target dispersion complement the theoretical correctness results. We therefore compared SIR, RQMC-SIR, random-walk Markov chain Monte Carlo (MCMC), and direct draws from a local Laplace proposal. The grid used $p\in\{3,4,5\}$, three centers, $\sigma\in\{0.25,0.75\}$, and 50 planned replications per cell. Each replication retained 2,000 samples, while SIR and RQMC-SIR used 32,768 proposals. MCMC used 2,000 burn-in iterations, thinning by five, and a chart-space scale of 0.8. The Laplace draws represent the uncorrected local approximation.

The three target-directed methods gave similar radial second moments. Their maximum difference in any row of Table~\ref{tab:sim4_radial} was 0.0062. For $p>2$, where exact sampling is unavailable, this agreement provides a useful cross-method check of target behavior. At $\sigma=0.75$, the Laplace moment was below SIR by 0.303, 0.763, and 1.485 for $p=3$, $p=4$, and $p=5$.

\begin{table}[ht]
\centering
\caption{Mean radial second moment $\rho(C,C_0)^2$, aggregated over three centers. Each fully completed method, dimension, and scale combination contains 150 replications. At $p=5$ and $\sigma=0.75$, SIR and RQMC-SIR contain 137 and 138 completed replications, respectively.}
\label{tab:sim4_radial}
\begin{tabular}{cccccc}
\toprule
$p$ & $\sigma$ & SIR & RQMC-SIR & MCMC & Laplace proposal \\
\midrule
3 & 0.25 & 0.1881 & 0.1884 & 0.1879 & 0.1836 \\
3 & 0.75 & 1.7266 & 1.7229 & 1.7197 & 1.4232 \\
4 & 0.25 & 0.3781 & 0.3783 & 0.3765 & 0.3637 \\
4 & 0.75 & 3.5456 & 3.5493 & 3.5446 & 2.7824 \\
5 & 0.25 & 0.6347 & 0.6336 & 0.6353 & 0.6061 \\
5 & 0.75 & 6.0779 & 6.0799 & 6.0737 & 4.5933 \\
\bottomrule
\end{tabular}
\end{table}

Average MCMC acceptance was between 0.54 and 0.56 at $p=3$, compared with 0.24 to 0.28 at $p=5$. Proposal-weight ESS was about 10,000 to 11,950 at $p=3$ and 2,530 to 4,280 at $p=5$, from 32,768 proposals. These trends identify dimension as the main factor governing finite-run efficiency in both Markov chain and importance-resampling implementations.

\subsection{Application 1: rolling financial correlations}\label{sec:finance_application}

The financial application asks whether quotient-affine geometry improves the description and held-out prediction of changing dependence across major asset classes. We obtained adjusted daily closing prices from Yahoo Finance through \texttt{yfinance} Python library \citep{aroussi_2025_YfinanceDownloadMarket}.  The four exchange-traded funds represent United States equities (SPY), developed-market equities outside the United States and Canada (EFA), intermediate United States Treasury bonds (IEF), and gold (GLD). The cached data contain 5,031 trading dates from January 3, 2006, through December 31, 2025. 

We computed correlation matrices from 60-trading-day windows of daily log returns. The inferential series advances by 60 trading days, so its return windows do not overlap. This produced 83 full-rank matrices, divided chronologically into 50 training, 12 validation, and 21 test observations using cutoffs at the end of 2017 and 2020. Their minimum eigenvalues ranged from 0.023 to 0.337, and their largest condition number was 98.9. A second series advances by 20 trading days and contains 249 matrices. It is used only for descriptive visualization because its windows overlap.

We fixed two components for all four mixture models. The proposed fit is an isotropic QA Fr\'echet mixture with weighted Fr\'echet updates and the leading local normalizer. This construction provides a computationally tractable approximation to mixture maximum likelihood. The baselines are spherical Gaussian mixture models (GMMs) fitted to raw off-diagonal entries, pairwise Fisher-transformed entries, and normalized-Cholesky coordinates. Every held-out log density was converted to Lebesgue measure on the six free off-diagonal correlation entries. The transformation Jacobian was included when the fitting coordinates differed from that measure, so the reported scores are comparable across models.

\begin{figure}[ht]
\centering
\includegraphics[width=0.98\textwidth]{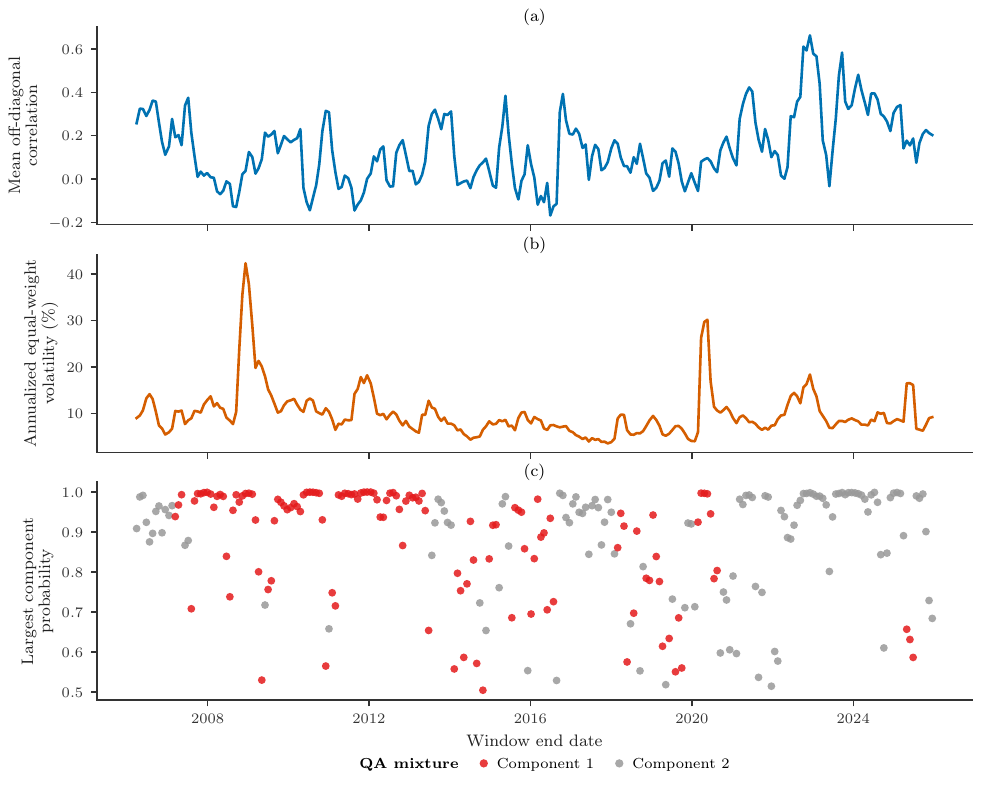}
\caption{Descriptive financial sequence based on 60-trading-day windows advanced by 20 days. (a) Mean off-diagonal correlation. (b) Annualized equal-weight realized volatility. (c) Largest QA mixture membership probability, colored by assigned component. These 249 overlapping windows are used only to visualize fitted dependence patterns over time.}
\label{fig:finance_timeline}
\end{figure}

Figure~\ref{fig:finance_timeline} places the fitted QA mixture in the temporal context of the denser descriptive series. Component 1 is assigned to 127 of the 249 overlapping windows and has mean annualized equal-weight volatility of 11.4\%. Component 2 is assigned to 122 windows and has mean volatility of 8.4\%. These overlapping windows are used only for description. They do not enter the held-out comparison or its uncertainty summaries.

Table~\ref{tab:finance_scores} reports the test-window comparison. The QA mixture has the largest mean held-out log density, followed by the normalized-Cholesky model. Its mean advantage over normalized Cholesky is 0.947 log-density units per window, and its score is larger in 16 of the 21 test windows. The corresponding mean advantages over Fisher and raw coordinates are 4.009 and 7.049. The nonoverlapping construction removes direct data reuse across adjacent matrices. Noncircular moving-block bootstrap intervals then provide dependence sensitivity summaries for the financial time series. These intervals use blocks of three windows, 50,000 resamples, and fixed method-specific offsets.

\begin{table}[t]
\centering
\small
\caption{Predictive log densities for 21 financial test windows. Score standard errors are across test windows. For each baseline, the difference is QA minus the named baseline, and its parenthetical value is a paired standard error. The last column is a 95\% noncircular moving-block bootstrap sensitivity interval for that difference. All scores use the same Lebesgue reference measure on free off-diagonal entries.}
\label{tab:finance_scores}
\begin{tabular}{cccc}
\toprule
Model & Mean score (SE) & QA difference (SE) & Block interval \\
\midrule
QA Fr\'echet mixture & \phantom{$-$}1.641 (0.503) & --- & --- \\
Normalized Cholesky & \phantom{$-$}0.694 (0.572) & 0.947 (0.285) & [0.502,\ 1.587] \\
Fisher coordinates & $-2.368$ (0.856) & 4.009 (0.725) & [2.855,\ 5.806] \\
Raw off-diagonals & $-5.408$ (1.526) & 7.049 (1.423) & [4.508,\ 10.833] \\
\bottomrule
\end{tabular}
\end{table}

\begin{figure}[ht]
\centering
\includegraphics[width=0.90\textwidth]{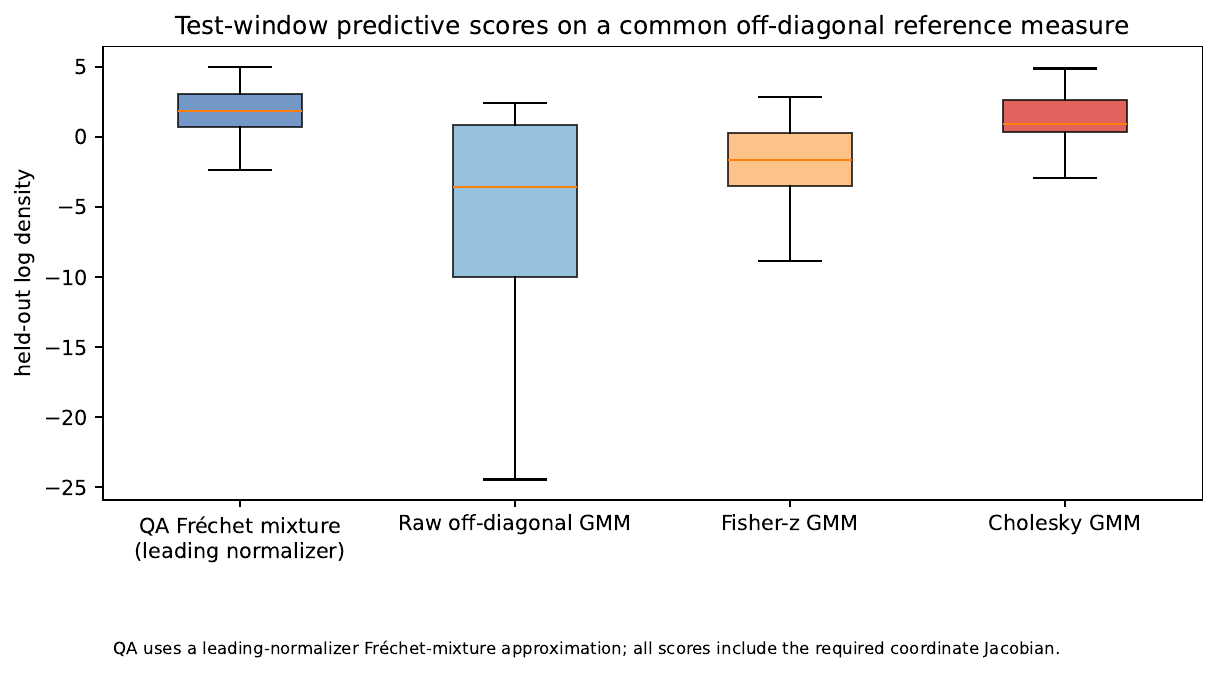}
\caption{Held-out log densities for the 21 nonoverlapping financial test windows. The QA model uses the leading-normalizer Fr\'echet-mixture approximation. Each coordinate model includes the change-of-variables term needed to express its density on the common off-diagonal reference measure.}
\label{fig:finance_scores}
\end{figure}

\begin{figure}[H]
\centering
\includegraphics[width=.92\textwidth]{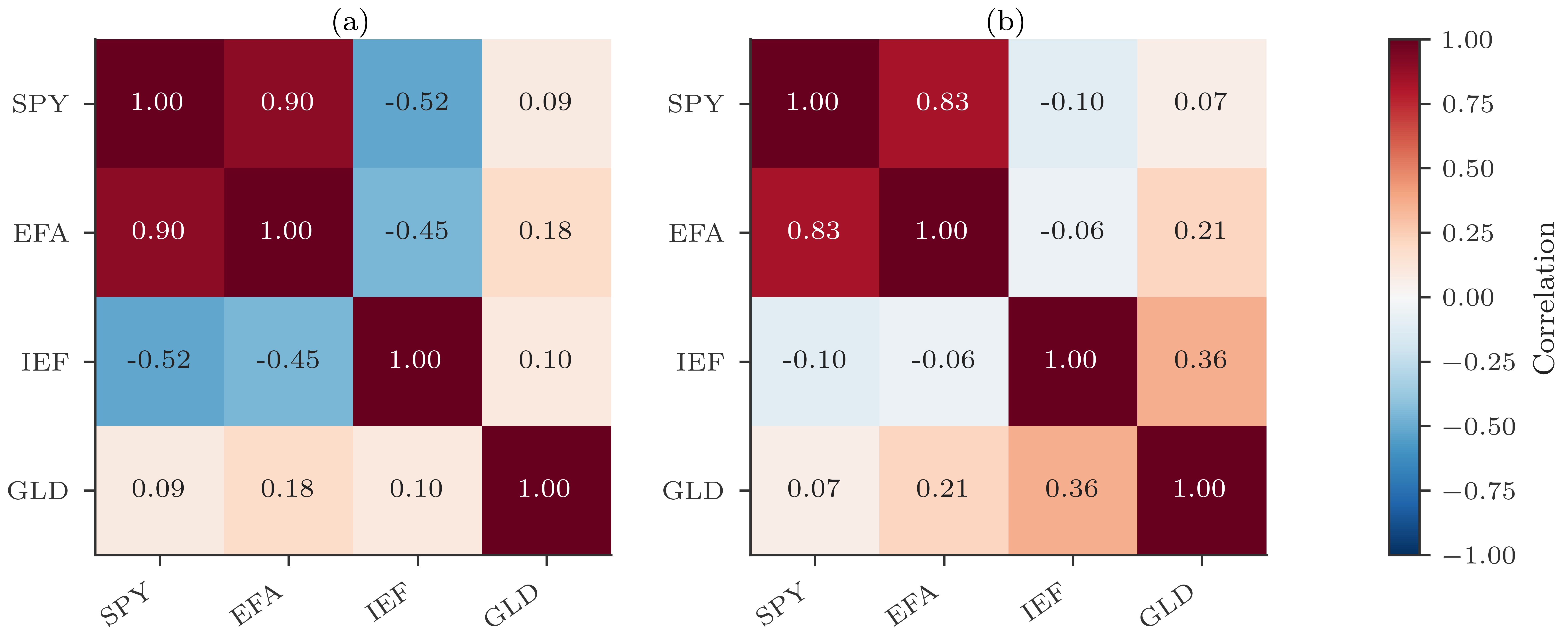}
\caption{Estimated centers of the two-component QA Fr\'echet mixture fitted to 50 nonoverlapping training windows. Component 1 in panel (a) has estimated weight 0.65 and Riemannian scale 0.35. Component 2 in panel (b) has estimated weight 0.35 and scale 0.34. Both matrices use the same color scale.}
\label{fig:finance_centers}
\end{figure}

The fitted QA centers reveal the main source of separation. Component 1 has mixing weight 0.65 and combines strong SPY--EFA dependence with correlations of $-0.52$ and $-0.45$ between IEF and the two equity funds. Component 2 has weight 0.35, retains strong SPY--EFA dependence, and places the equity--IEF correlations near zero. Its IEF--GLD correlation is 0.36, compared with 0.10 in Component 1. Figure~\ref{fig:finance_centers} displays these centers. We describe the components through their estimated dependence patterns rather than assigning unobserved labels such as ``calm'' or ``stress.''

The QA mixture fit required 1,165 seconds, whereas each coordinate GMM required about 0.4 seconds on the same machine. This difference quantifies the computational cost of fitting the mixture directly on quotient geometry. Within the controlled setting of four assets and 21 test matrices, the held-out results support the reported density advantage.

\subsection{Application 2: a centered prior for a latent correlation matrix}\label{sec:lkj_application}

The second application examines how an externally supplied QA center affects inference when it is aligned with or displaced from the true latent correlation matrix. We use a controlled synthetic study, which permits direct evaluation of recovery and held-out prediction against known truth. For group $g$, a latent random effect has covariance $D C_\star D$, where $D=\diag(0.8,1.0,1.2)$ and $C_\star$ is the true $3\times3$ correlation matrix. Each observed group mean averages four replicates with residual standard deviation 0.5. Every replication contains 30 training groups and 200 held-out groups.

Four scenarios separate center information from boundary difficulty. The identity scenario uses the identity as both truth and QA center. The matched-block scenario sets correlation 0.55 between variables 1 and 2 for both matrices. Under center misspecification, the true 0.55 correlation instead links variables 2 and 3, while the QA template retains variables 1 and 2. The near-boundary scenario uses a true correlation of 0.85 and a QA template correlation of 0.55 between variables 1 and 2.

The QA prior has Riemannian scale 0.5, while the LKJ prior has shape parameter 2. Both posterior targets use the normalized-Cholesky chart with their required Jacobians. The fixed-hyperparameter design makes their normalizing constants cancel from posterior ratios and focuses the comparison on posterior recovery and prediction. Each scenario has 50 replications. We ran four adaptive random-walk chains per prior with 800 warmup iterations and 1,000 retained draws. 

Table~\ref{tab:lkj_comparison} reports paired QA minus LKJ differences. The QA prior reduces center error and entrywise root mean squared error in the identity, matched-block, and near-boundary scenarios. It also improves held-out prediction in these scenarios. The misspecified-center scenario quantifies sensitivity to the supplied information. There, QA center error increases by 0.024 and held-out log score decreases by 0.005 per group.

\begin{table}[ht]
\centering
\footnotesize
\setlength{\tabcolsep}{4pt}
\caption{Controlled prior comparison over 50 paired replications per scenario. The first three columns are QA minus LKJ, with paired Monte Carlo standard errors in parentheses. Negative error differences and positive held-out score differences favor QA. Coverage entries are corrected means with standard errors across replications for QA/LKJ, based on nominal 90\% marginal intervals for the three off-diagonal correlations.}
\label{tab:lkj_comparison}
\begin{tabular}{ccccc}
\toprule
Scenario & Center error & Off-diagonal RMSE & Held-out score & Coverage QA/LKJ \\
\midrule
Identity & $-0.035$ (0.004) & $-0.013$ (0.002) & \phantom{$-$}0.009 (0.002) & 0.927 (0.022) / 0.920 (0.022) \\
Matched block & $-0.087$ (0.012) & $-0.032$ (0.004) & \phantom{$-$}0.016 (0.003) & 0.940 (0.018) / 0.907 (0.025) \\
Misspecified center & \phantom{$-$}0.024 (0.017) & \phantom{$-$}0.014 (0.006) & $-0.005$ (0.004) & 0.853 (0.029) / 0.927 (0.020) \\
Near boundary & $-0.073$ (0.009) & $-0.012$ (0.001) & \phantom{$-$}0.011 (0.002) & 0.900 (0.024) / 0.907 (0.023) \\
\bottomrule
\end{tabular}
\end{table}

\begin{figure}[ht]
\centering
\includegraphics[width=0.95\textwidth]{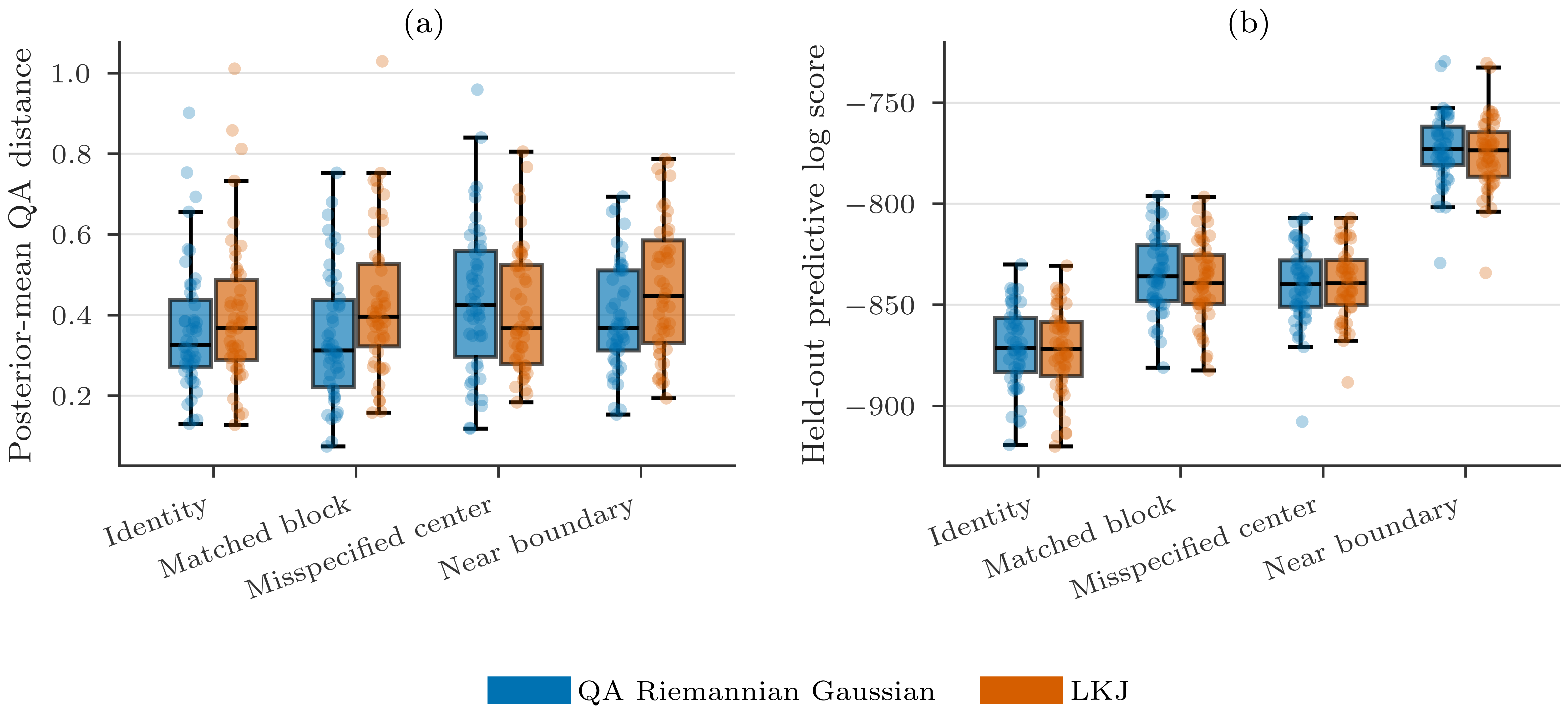}
\caption{Posterior comparison for the QA and LKJ priors over 50 replications per scenario. (a) Quotient-affine distance from the posterior mean to the truth, for which smaller values are better. (b) Predictive log density summed over 200 held-out groups, for which larger values are better. QA gains appear in three scenarios, while the misspecified scenario displays sensitivity to center quality.}
\label{fig:lkj_performance}
\end{figure}

Coverage was computed from the strict-lower-triangle entries of the archived posterior draws, pooling 4,000 draws per replication and prior. It is near 0.90 in three scenarios and reaches 0.853 under center misspecification. Figure~\ref{fig:lkj_performance} shows the replication-level patterns behind the average differences.

\begin{figure}[ht]
\centering
\includegraphics[width=\textwidth]{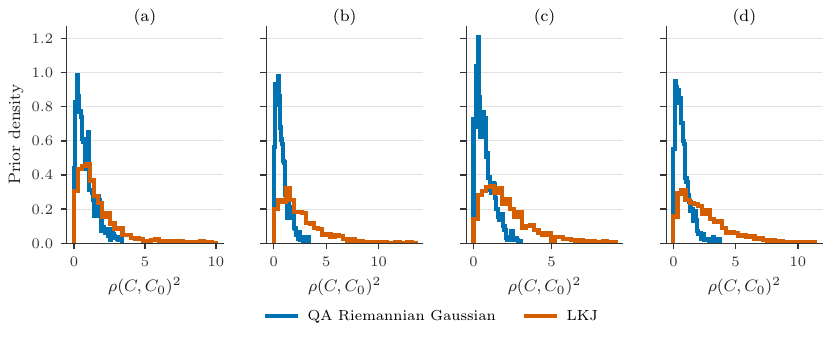}
\caption{Prior squared QA distance from the supplied template under the fixed hyperparameters used in the posterior study. Panels (a) through (d) correspond to the identity, matched-block, misspecified-center, and near-boundary scenarios. Each density pools 4,000 prior draws. The displayed fixed hyperparameters yield different radial concentrations for the QA and LKJ priors.}
\label{fig:lkj_prior_spread}
\end{figure}

The fixed QA and LKJ priors also differ in radial concentration. Across scenarios, the QA prior's mean squared radial distance ranges from 0.74 to 0.86, compared with 1.71 to 2.65 for LKJ. The differences in Table~\ref{tab:lkj_comparison} therefore summarize the combined effects of prior geometry, center information, and concentration. Figure~\ref{fig:lkj_prior_spread} displays the prior-distance distributions and documents the tighter concentration of QA around its supplied template.

No chain generated an invalid correlation proposal, and no QA distance evaluation remained unresolved. Across fits, 61.8\% had maximum split $\widehat R\leq1.02$, while 97.8\% were below 1.05. Average minimum approximate effective sample sizes ranged from 190 to 307, with the smallest values near the boundary. These diagnostics characterize the Monte Carlo precision of the finite-run comparison and identify the near-boundary setting as the most demanding.

A four-chain QA fit required 30 to 36 seconds on average, compared with 1.0 to 1.3 seconds for LKJ. The resulting ratios ranged from 26 to 30. The practical gains are strongest when the supplied center is informative. Prior concentration, Monte Carlo precision, and computation time remain relevant design choices when applying the QA prior.

\section{Discussion}

We proposed a normalized Riemannian Gaussian model for full-rank correlation matrices under quotient-affine geometry. To our knowledge, this is the first likelihood model on the open elliptope that combines quotient-affine distance with the corresponding Riemannian volume while explicitly accounting for the normalizing constant. The construction provides a statistical model in which marginal scales are removed by design, rather than through a particular coordinate representation. A central implication is that, beyond the two-dimensional Fisher-transform case, the model center and the population \Frechet mean need not coincide because normalization may vary across the manifold.

The theoretical existence of this difference does not, however, determine its practical importance. In the estimation experiments, the tested likelihood corrections changed center and scale recovery only slightly, whereas importance-based normalization greatly increased computation. This suggests that future work should characterize not only whether $Z(\bar C,\sigma)$ depends on $\bar C$, but also whether its variation is large relative to sampling uncertainty, integration error, and the inferential quantity of interest. Finite-dispersion approximations and centerwise sensitivity measures for $\log Z$ would help identify regimes in which \Frechet or leading-Laplace fitting is adequate and those in which normalized likelihood can materially alter estimation. Such results would provide a more useful criterion for choosing an estimation method than center dependence alone.

The experiments also exposed computational difficulties that become more pronounced with dimension and dispersion. Importance-sampling ESS declined, several high-dispersion runs did not complete, and uncorrected local Laplace draws became increasingly underdispersed. These patterns indicate that proposals based primarily on local geometry may fail to represent probability mass extending toward difficult regions of the elliptope. They also interact with two lower-level bottlenecks: the quotient distance requires an inner optimization that is not globally certified, and direct evaluation of the chart-volume density scales poorly with dimension. Future computational work should therefore combine more stable quotient derivatives and stronger stationarity diagnostics with globally adapted, heavy-tailed, or manifold-based proposals \citep{girolami_2011_RiemannManifoldLangevin,byrne_2013_GeodesicMonteCarlo}. It would also be useful to propagate both distance-optimization error and normalizer-estimation error into likelihood-based uncertainty. Until such diagnostics are available, ``exact likelihood'' should continue to refer to the statistical target rather than to numerically exact evaluation.

The applications raise related modeling questions. In the financial example, the QA mixture improved held-out density under a common reference measure, but the fitted model used the leading-normalizer \Frechet approximation and a static two-component specification. Component-specific normalization could change estimated weights and centers, while serial dependence among rolling correlations is not represented by an independent mixture. Dynamic, state-space, or composite-likelihood extensions would provide a more direct treatment of evolving dependence and would allow assessment across different assets, window lengths, and numbers of components. In the prior study, a substantively credible center improved recovery, whereas a misspecified center could be harmful. Moreover, the QA and LKJ priors were not matched in radial concentration. Future comparisons should therefore calibrate prior spread before separating geometric effects from differences in informativeness. Hierarchical, robust, or mixture distributions for the center would also allow uncertainty about an external template to enter the model rather than being treated as fixed.

Several theoretical extensions would support these computational and applied developments. Center dependence was established only for $p=3$ in the small-dispersion regime, so the behavior of $Z$ at finite dispersion and in higher dimensions remains to be characterized. Likelihood asymptotics beyond compact-set consistency would require local smoothness of quotient logarithms, control near the cut locus, and nonsingularity of the information matrix. The present isotropic model also imposes a single radial scale and applies only to full-rank correlation matrices. Anisotropic or wrapped constructions may offer greater flexibility, but would require tractable normalizers and, for wrapped models, exponential-map Jacobians and treatment of multiple preimages \citep{thanwerdas_2021_GeodesicQuotientAffineMetrics}. Rank-deficient correlation matrices require a separate geometric formulation.

Overall, the proposed model is most immediately useful when quotient invariance is scientifically essential, a meaningful correlation center is available, and dimension and dispersion remain moderate. In that setting, it supplies an intrinsic likelihood and a coherent sampling target. The limitations observed here also delineate the next questions: determining when normalization has inferential consequence, developing algorithms that remain stable away from local regimes, and extending the model to richer dependence structures without losing its quotient-geometric interpretation.

\clearpage
\spacingset{1}

\phantomsection
\label{supplementary-material}
\begin{center}
  {\LARGE\bfseries Supplementary Materials\par}
\end{center}
\bigskip

\appendix
\setcounter{figure}{0}
\setcounter{table}{0}
\setcounter{equation}{0}

\renewcommand{\thesection}{S\arabic{section}}
\renewcommand{\thesubsection}{S\arabic{section}.\arabic{subsection}}
\renewcommand{\thefigure}{S\arabic{figure}}
\renewcommand{\thetable}{S\arabic{table}}
\renewcommand{\theequation}{S\arabic{equation}}

\renewcommand{\theHsection}{supp.\arabic{section}}
\renewcommand{\theHsubsection}{supp.\arabic{section}.\arabic{subsection}}
\renewcommand{\theHfigure}{supp.figure.\arabic{figure}}
\renewcommand{\theHtable}{supp.table.\arabic{table}}
\renewcommand{\theHequation}{supp.equation.\arabic{equation}}

\noindent
This supplement contains the failure audit and reproducibility details for the computational studies, followed by all proofs for the main paper. The proof section also verifies the global normalized-Cholesky chart used for computation.

\section{Computational diagnostics and reproducibility}\label{sec:supp_failures}

Completed-case summaries can conceal numerical instability, so failures require separate accounting. Table~\ref{tab:completion} identifies the stress settings that challenge the full-rank implementation. Every recorded failure was a numerical validation error at a proposal or generated matrix whose minimum eigenvalue lay between approximately $10^{-13}$ and $10^{-10}$. Failures were concentrated in the largest-dimension, largest-dispersion cells. They do not imply that the mathematical target contains indefinite matrices. Rather, near-boundary full-rank draws can fall below the implementation's numerical rank threshold. Failed rows were excluded from numerical means, so incomplete stress-cell summaries are selection-conditioned.

\begin{table}[H]
\centering
\caption{Exception-level completion of the manuscript-scale run. The top-level audit intentionally returned a nonzero status because at least one job failed, even though all runner and plotting stages completed.}
\label{tab:completion}
\begin{tabular}{lrrr}
\toprule
Study & Completed & Planned & Failed \\
\midrule
Simulation 1 & 1,000 & 1,000 & 0 \\
Simulation 2 & 1,960 & 2,016 & 56 \\
Simulation 3, main grid & 9,576 & 9,600 & 24 \\
Simulation 3, importance subset & 540 & 540 & 0 \\
Simulation 4 & 3,575 & 3,600 & 25 \\
\midrule
Total & 16,651 & 16,756 & 105 \\
\bottomrule
\end{tabular}
\end{table}

The reproducibility materials contain the source code, command-line runners, seeds, JSON configuration files, raw comma-separated result files, the timestamped execution log, and vector and raster versions of every generated figure. The top-level script fixes numerical linear-algebra libraries to one thread per process and dispatches one independent job per worker. It audits every raw result file and returns a nonzero status when a row has \texttt{success=False}. The reported results precede any rerun under a revised numerical policy. Retaining their failure counts records an important limitation of the current implementation.

\section{Proofs}\label{sec:supp_proofs}

This section collects all arguments supporting claims in the main paper. The proofs follow the order of the main text, except where a later result supplies a needed identity. References to statements and equations use the numbering established in the preceding text.

\subsection{Verification of the normalized-Cholesky chart}\label{sec:supp_chart}

The computational chart must cover every full-rank correlation matrix without ambiguity. Let $C=L L^T$ be the positive-diagonal Cholesky factorization and set $\Gamma=\Theta(C)$. The Cholesky factor in \eqref{eq:cholesky_chart_inverse} is $\Diag(\Gamma\Gamma^T)^{-1/2}\Gamma$, so applying $\Theta$ returns $\Gamma$. The Cholesky factorization and the displayed maps are smooth. Hence the chart is a smooth bijection with smooth inverse.

\subsection{Proof of Lemma~\ref{lem:horizontal_lift}}

The quotient metric requires a horizontal ambient representative for each tangent direction on $\Corr_p$. We first characterize horizontality and then solve for the unique diagonal correction.

The vertical space at $C$ is $\mathcal V_C=\{AC+CA:A\text{ diagonal}\}$. A vector $U\in T_C\SPD^p$ is horizontal if it is orthogonal to every $AC+CA$ under $g^{\AI}$. For diagonal $A$,
\[
g_C^{\AI}(U,AC+CA)=\tr(C^{-1}U C^{-1}AC)+\tr(C^{-1}UC^{-1}CA).
\]
Using cyclic invariance of the trace, this equals
\[
2\tr\{A C^{-1}U\}=2a^T\diag(C^{-1}U),
\]
where $a$ is the diagonal of $A$. Thus $U$ is horizontal if and only if $\diag(C^{-1}U)=0$.

Let $U=H+AC+CA$. Since $\diag(H)=0$, $C$ has unit diagonal, and $\diag(AC+CA)=2a$, formula~\eqref{eq:dpi_formula} gives $d\pi_C(U)=H$. It remains to choose $a$ so that $U$ is horizontal. We have
\[
\diag\{C^{-1}(H+AC+CA)\}=\diag(C^{-1}H)+a+\diag(C^{-1}AC).
\]
The $i$th component of $\diag(C^{-1}AC)$ is $\sum_j(C^{-1})_{ij}a_jC_{ji}$, which is the $i$th component of $B(C)a$. Hence the horizontality condition is
\[
\{I+B(C)\}a=-\diag(C^{-1}H).
\]
The matrix $B(C)=C^{-1}\Had C$ is positive semidefinite by the Schur product theorem because both $C^{-1}$ and $C$ are positive definite. Therefore $I+B(C)$ is positive definite and the solution is unique. This proves the lemma. \qed

\subsection{Proof of Proposition~\ref{prop:geometric_foundation}}

The model requires a complete smooth quotient with attained distances and controlled volume growth. We establish these properties from the diagonal action, Riemannian submersion, and curvature comparison.

The action is free because $D\Sigma D=\Sigma$ implies $d_i^2\Sigma_{ii}=\Sigma_{ii}$ for every $i$; since $\Sigma_{ii}>0$ and $d_i>0$, $d_i=1$. It is proper as follows. If $\Sigma_n\to\Sigma$ and $D_n\Sigma_nD_n\to\Lambda$ in $\SPD^p$, then the diagonal entries give $d_{n,i}^2(\Sigma_n)_{ii}\to\Lambda_{ii}$, so $d_{n,i}\to(\Lambda_{ii}/\Sigma_{ii})^{1/2}\in(0,\infty)$. Hence $D_n$ has a convergent subsequence in $\Diagp(p)$. The action is isometric because the affine-invariant metric is invariant under congruence.

A free and proper action gives a smooth quotient manifold by the quotient manifold theorem \citep{lee_2012_IntroductionSmoothManifolds}. Since the action is isometric, the quotient inherits the metric that makes the projection a Riemannian submersion \citep{lee_1997_RiemannianManifoldsIntroduction}.

The quotient is geodesically complete. A base geodesic has a horizontal lift. For an isometric action, the fundamental vertical fields are Killing fields, and their inner products with a total-space geodesic velocity are conserved. An initially horizontal lift therefore remains horizontal and projects to the base geodesic while the total-space geodesic exists \citep{oneill_1966_FundamentalEquationsSubmersion}. Completeness of $(\SPD^p,g^{\AI})$ gives existence for all time.

By Hopf--Rinow, closed bounded subsets of the quotient are compact \citep{sakai_1996_RiemannianGeometry}. The orbit $\{DC_2D:D\in\Diagp(p)\}$ is closed by properness, so its distance from $C_1$ is attained. This proves attainment in \eqref{eq:qa_distance}. Horizontal persistence also justifies reading \eqref{eq:qa_log_definition} as the initial velocity of a minimizing quotient geodesic when the representative is unique.

On a complete smooth Riemannian manifold, the cut locus of a point has Riemannian volume zero and squared distance from that point is smooth outside the point and its cut locus \citep{sakai_1996_RiemannianGeometry}. This gives part (c).

The affine-invariant SPD space is a nonpositively curved symmetric space with sectional curvature bounded below by $-\kappa_p$ for some finite $\kappa_p$. O'Neill's curvature formula for a Riemannian submersion gives, on horizontal two-planes \citep{oneill_1966_FundamentalEquationsSubmersion},
\[
\sec_{\QA}\ge \sec_{\AI}\ge -\kappa_p.
\]
Therefore $\Ric_{\QA}\ge-(d-1)\kappa_p$. Bishop--Gromov comparison gives an exponential upper bound on ball volumes, uniform in the center, of the form in \eqref{eq:volume_growth} \citet[Chapter~IV]{sakai_1996_RiemannianGeometry}. Let $V_{\bar C}(r)=\vol_{\QA}\{B_{\QA}(\bar C,r)\}$ and $W(r)=A_p e^{b_pr}$. For any nonnegative decreasing function $f$, layer-cake gives
\[
\int f(r)\,dV_{\bar C}(r)=\int_0^\infty V_{\bar C}\{r:f(r)>t\}\,dt\le \int_0^\infty W\{r:f(r)>t\}\,dt=\int f(r)\,dW(r).
\]
Applying this with $f(r)=\exp\{-r^2/(2\sigma^2)\}$ gives finiteness of the unweighted Gaussian radial integral. For moments, use $r^k e^{-r^2/(2\sigma^2)}\le c_{k,\sigma} e^{-r^2/(4\sigma^2)}$ and apply the same decreasing-function bound. Thus all Gaussian-weighted radial moments in part (d) are finite. \qed

\subsection{Proof of Proposition~\ref{prop:well_defined}}

The quotient density must not depend on the chosen covariance representatives. Congruence invariance reduces this claim to a reparameterization of the positive diagonal optimizer.

Let $\Sigma'=A\Sigma A$ and $\bar\Sigma'=B\bar\Sigma B$ with $A,B\in\Diagp(p)$. By congruence invariance of $d_{\AI}$,
\[
d_{\AI}(\Sigma',D\bar\Sigma'D)=d_{\AI}(\Sigma,A^{-1}DB\bar\Sigma BDA^{-1}).
\]
As $D$ ranges over $\Diagp(p)$, so does $A^{-1}DB$. Thus the infimum is unchanged. The quotient volume measure is defined by the Riemannian submersion construction from Proposition~\ref{prop:geometric_foundation}, so the density is independent of representatives. \qed

\subsection{Proof of Proposition~\ref{prop:coordinate_density}}

Numerical work uses the normalized-Cholesky chart, so the invariant density must be expressed relative to Lebesgue measure. This follows directly from the coordinate form of Riemannian volume.

The Riemannian volume form in coordinates is $\sqrt{\det G(x)}\,dx=J_{\QA}(x)dx$. Substituting the chart map into Definition~\ref{def:rg} gives \eqref{eq:coordinate_density}. Integrating that equation over $\R^d$ gives \eqref{eq:chart_integral_Z}. \qed

\subsection{Proof of Proposition~\ref{prop:max_entropy}}

The maximum-entropy claim follows by comparing any feasible density with the proposed radial density through Kullback--Leibler divergence.

Let $p_0(C)=Z^{-1}\exp\{-\beta\rho(C,\bar C)^2\}$ with $\beta=1/(2\sigma^2)$. For any feasible density $q$,
\[
0\le \int q\log\frac{q}{p_0}\,d\vol_{\QA}=\int q\log q\,d\vol_{\QA}+\log Z+\beta m.
\]
The last two terms are fixed over the feasible class. Hence $-\int q\log q\,d\vol_{\QA}$ is maximized by $q=p_0$, with equality if and only if $q=p_0$ almost everywhere. \qed

\subsection{Proof of Proposition~\ref{prop:p2}}

The analytic $p=2$ form follows from the quotient metric along the single correlation coordinate. Fisher's transformation then makes both the metric and volume element constant.

Let $C=C(r)$ and let $H=h\begin{pmatrix}0&1\\1&0\end{pmatrix}$. The horizontal lift has the form
\[
U=\frac{h}{1-r^2}\begin{pmatrix}r&1\\1&r\end{pmatrix},
\]
which satisfies $d\pi_C(U)=H$ and $\diag(C^{-1}U)=0$. Direct calculation gives
\[
g_C^{\QA}(H,H)=\tr(C^{-1}UC^{-1}U)=\frac{2h^2}{(1-r^2)^2}.
\]
Thus the line element is $ds^2=2(1-r^2)^{-2}dr^2$. With $z=\operatorname{atanh}(r)$, $dz=dr/(1-r^2)$, so $ds^2=2dz^2$ and $d\vol_{\QA}=\sqrt2\,dz$. The geodesic distance is therefore \eqref{eq:p2_distance}. The normalizing constant is
\[
Z=\sqrt2\int_{-\infty}^{\infty}\exp\left\{-\frac{2(z-z_0)^2}{2\sigma^2}\right\}dz=\sqrt{2\pi}\,\sigma.
\]
This proves the proposition. \qed

\subsection{Proof of Corollary~\ref{cor:p2_mle}}

Proposition~\ref{prop:p2} reduces the model to an ordinary normal likelihood in Fisher coordinates. The stated estimators then follow from the usual Gaussian MLE.

By Proposition~\ref{prop:p2}, the density in $z$ coordinates is normal with mean $z_0$ and variance $\sigma^2/2$. The usual normal likelihood gives the MLEs in \eqref{eq:p2_mle}; $\widehat r_0=\tanh(\widehat z_0)$ follows by invariance of the MLE under smooth reparametrization. \qed

\subsection{Proof of Proposition~\ref{prop:center_dependence}}

To prove center dependence, it is enough to show that scalar curvature varies along a tractable family and then apply Proposition~\ref{prop:laplace}. We compute that curvature in local correlation coordinates.

Scalar curvature is coordinate invariant, so it is convenient to compute it in the local correlation coordinates
\[
C(r,s,t)=\begin{pmatrix}1&r&s\\ r&1&t\\ s&t&1\end{pmatrix}
\]
near the line $(r,s,t)=(r,0,0)$, where $|r|<1$. This line is the same family as in the proposition, with
\[
r=\frac{a}{\sqrt{1+a^2}},\qquad q=1-r^2.
\]
For the coordinate tangent matrices $E_r,E_s,E_t$, we apply Lemma~\ref{lem:horizontal_lift} and form
\[
G_{ij}=\tr(C^{-1}E_i^{\mathrm H}C^{-1}E_j^{\mathrm H}),\qquad i,j\in\{r,s,t\}.
\]
Only the second-order Taylor expansion in $(s,t)$ is needed for scalar curvature on the line. Direct substitution in Lemma~\ref{lem:horizontal_lift} gives
\begin{align*}
G_{rr}&=\frac{2}{q^2}+\frac{(5r^2+4)(s^2+t^2)-2r(r^2+8)st}{2q^3}+O(\|(s,t)\|^3),\\
G_{rs}&=\frac{3rs-(r^2+2)t}{q^2}+O(\|(s,t)\|^3),\qquad
G_{rt}=\frac{3rt-(r^2+2)s}{q^2}+O(\|(s,t)\|^3),\\
G_{ss}&=\frac{2}{q}+\frac{(r^2+8)s^2+(r^2+2)^2t^2-2r(r^2+8)st}{2q^2}+O(\|(s,t)\|^3),\\
G_{tt}&=\frac{2}{q}+\frac{(r^2+2)^2s^2+(r^2+8)t^2-2r(r^2+8)st}{2q^2}+O(\|(s,t)\|^3),\\
G_{st}&=-\frac{2r}{q}+\frac{\{-r^4+13r^2+6\}st+r(r^2-10)(s^2+t^2)}{2q^2}+O(\|(s,t)\|^3).
\end{align*}
These expressions follow algebraically from the linear systems in Lemma~\ref{lem:horizontal_lift}.

The omitted terms in $G_{rs}$ and $G_{rt}$ require a symmetry argument. Congruence by $\operatorname{Diag}(1,1,-1)$ induces the quotient isometry $(r,s,t)\mapsto(r,-s,-t)$ and fixes the line $s=t=0$. The metric pullback identity makes $G_{rs}$ and $G_{rt}$ odd in $(s,t)$. Their even-order Taylor coefficients therefore vanish, so there are no quadratic terms. The displayed expansion contains every metric derivative needed for the Christoffel symbols and their first derivatives on the line.

At $s=t=0$, the metric and its inverse are
\[
G=\begin{pmatrix}
2/q^2&0&0\\
0&2/q&-2r/q\\
0&-2r/q&2/q
\end{pmatrix},\qquad
G^{-1}=\begin{pmatrix}
q^2/2&0&0\\
0&1/2&r/2\\
0&r/2&1/2
\end{pmatrix}.
\]
Using
\[
\Gamma^k_{ij}=\frac12G^{k\ell}(\partial_iG_{j\ell}+\partial_jG_{i\ell}-\partial_\ell G_{ij})
\]
and differentiating the displayed second-order expansion gives the Ricci tensor on the line:
\[
\Ric=\begin{pmatrix}
-1/(2q^2)&0&0\\[2mm]
0&(3r^2-2)/(4q)&r(2-3r^2)/(4q)\\[2mm]
0&r(2-3r^2)/(4q)&(3r^2-2)/(4q)
\end{pmatrix}.
\]
Contracting with $G^{-1}$ yields
\[
\Scal=\tr(G^{-1}\Ric)=\frac{3(r^2-1)}4=-\frac34(1-r^2).
\]
Since $1-r^2=(1+a^2)^{-1}$, this proves \eqref{eq:scalar_curve}. The accompanying script \path{verify_p3_scalar_curvature.py} constructs the same metric tensor and checks the parity relation. It also validates the finite-difference routine on hyperbolic three-space and evaluates scalar curvature along the block-correlation curve.

Since $\Scal(C_0)\ne\Scal(C_1)$, Proposition~\ref{prop:laplace} implies
\[
Z(C_0,\sigma)-Z(C_1,\sigma)=(2\pi\sigma^2)^{3/2}\left\{\frac{\Scal(C_1)-\Scal(C_0)}{6}\sigma^2+O(\sigma^4)\right\},
\]
which is nonzero for all sufficiently small $\sigma>0$. \qed

\subsection{Proof of Corollary~\ref{cor:frechet_drift}}

The corollary relates the gradient of the population Fr\'echet objective to the center derivative of $\log Z$. The score identity supplies this relation.

For points outside the cut locus, $\nabla_Q\rho(C,Q)^2=-2\Log^{\QA}_Q(C)$. The domination argument used in the proof of Theorem~\ref{thm:score}, based on Proposition~\ref{prop:geometric_foundation}, justifies interchanging the gradient and expectation in a neighborhood of $C_0$. Therefore
\[
\nabla F(C_0)=-2\E_{C_0,\sigma_0}\{\Log^{\QA}_{C_0}(C)\}.
\]
By Theorem~\ref{thm:score}, evaluated at the true parameter,
\[
\E_{C_0,\sigma_0}\{\Log^{\QA}_{C_0}(C)\}=\sigma_0^2\nabla_{C_0}\log Z(C_0,\sigma_0).
\]
If the right-hand side is nonzero, $\nabla F(C_0)\ne0$, so $C_0$ is not a stationary point of $F$ and cannot be its unique minimizer. Consistency of exact MLEs on compact sets follows from Theorem~\ref{thm:mle_consistency}. \qed

\subsection{Proof of Theorem~\ref{thm:score}}

The score equations require differentiation of $Z(\bar C,\sigma)$ under the integral. Proposition~\ref{prop:geometric_foundation} provides the domination needed for this step.

Outside the cut locus, $\nabla_{\bar C}\rho(C,\bar C)^2=-2\Log^{\QA}_{\bar C}(C)$. Proposition~\ref{prop:geometric_foundation} gives finite radial moments. For $\bar C$ in a compact neighborhood of $\bar C_0$ and $\sigma$ in a compact interval $[\sigma_{\min},\sigma_{\max}]\subset(0,\infty)$, let $R$ be an upper bound for $\rho(\bar C,\bar C_0)$ over that neighborhood. The gradient of the kernel is dominated by a constant multiple of
\[
(1+\rho(C,\bar C_0))\exp\left\{-\frac{(\rho(C,\bar C_0)-R)_+^2}{2\sigma_{\max}^2}\right\},
\]
which is integrable by the exponential volume-growth bound. Thus differentiation under the integral is justified by dominated convergence. We obtain
\[
\nabla_{\bar C}Z(\bar C,\sigma)=\frac{1}{\sigma^2}\int \Log^{\QA}_{\bar C}(C)\exp\left\{-\frac{\rho(C,\bar C)^2}{2\sigma^2}\right\}d\vol_{\QA}(C),
\]
which equals $Z(\bar C,\sigma)\sigma^{-2}\E_{\bar C,\sigma}\{\Log^{\QA}_{\bar C}(C)\}$. Differentiating \eqref{eq:loglik} gives \eqref{eq:center_score}. The scale derivative follows from
\[
\frac{\partial}{\partial\sigma}Z(\bar C,\sigma)=\frac{1}{\sigma^3}\int\rho(C,\bar C)^2\exp\left\{-\frac{\rho(C,\bar C)^2}{2\sigma^2}\right\}d\vol_{\QA}(C).
\]
This gives \eqref{eq:scale_score}. \qed

\subsection{Proof of Proposition~\ref{prop:profile_scale}}

The scale equation becomes transparent after writing the radial family as a one-parameter exponential family. Differentiating its second moment yields the required monotonicity.

For fixed $\bar C$, write $H(C)=\rho(C,\bar C)^2$ and $\beta=1/(2\sigma^2)$. The density is an exponential family in natural parameter $-\beta$ with sufficient statistic $H$. Standard differentiation gives
\[
\frac{\partial}{\partial \beta}\E_{\beta}[H]=-\Var_{\beta}(H).
\]
Since $d\beta/d\sigma=-\sigma^{-3}$, \eqref{eq:m2_monotone} follows. The scale equation is exactly \eqref{eq:scale_score_eq}. \qed

\subsection{Proof of Proposition~\ref{prop:identifiability}}

Identifiability can be checked in two stages. The unique density maximizer determines the center, and the radial decay then determines the scale.

The density $p(C\mid C_0,\sigma)$ is continuous everywhere, because $\rho$ is continuous, and it has its unique maximum at $C=C_0$, because $\rho(C,C_0)^2$ is zero only at $C_0$ and positive elsewhere. The Riemannian volume measure has full support. Hence if two continuous densities are equal almost everywhere, they are equal everywhere. Equality of the unique maximizers gives $C_1=C_2$. With the center fixed, equality of densities implies
\[
-\frac{\rho(C,C_1)^2}{2\sigma_1^2}-\log Z(C_1,\sigma_1)=-\frac{\rho(C,C_1)^2}{2\sigma_2^2}-\log Z(C_1,\sigma_2)
\]
for all $C$ outside a null set. Since $\rho(C,C_1)^2$ is not constant, $\sigma_1=\sigma_2$. \qed

\subsection{Proof of Theorem~\ref{thm:mle_consistency}}

Consistency follows from compact-parameter M-estimation once identifiability, continuity, and an integrable envelope are established. The proof verifies these conditions for the exact likelihood.

Let $\theta=(Q,\sigma)\in\Theta$ and $\ell_1(\theta;C)=\log p(C\mid Q,\sigma)$. Identifiability follows from Proposition~\ref{prop:identifiability}. The expected log likelihood is therefore uniquely maximized at $(C_0,\sigma_0)$ by strict nonnegativity of Kullback--Leibler divergence. Continuity in $\theta$ follows from continuity of $\rho$ and dominated convergence for $Z$.

For any fixed $Q_*\in K$, the triangle inequality gives
\[
\rho(C,Q)^2\le 2\rho(C,Q_*)^2+2\operatorname{diam}(K)^2.
\]
Together with $\sigma\ge\sigma_{\min}$, compactness of $K$, and Proposition~\ref{prop:geometric_foundation}, this gives an integrable envelope for $\sup_{\theta\in\Theta}|\ell_1(\theta;C)|$.

The uniform law of large numbers and Wald's argmax theorem now apply. The relevant consistency results for M-estimators are given in  \citet[Chapter~5]{vandervaart_1998_AsymptoticStatistics}. This proves consistency. \qed

\subsection{Proof of Proposition~\ref{prop:laplace}}

The small-dispersion expansion is local around $\bar C$. We therefore separate the normal-coordinate contribution from the exponentially small contribution outside a fixed geodesic ball.

Choose $r_0$ smaller than the injectivity radius at $\bar C$ and split the integral into the geodesic ball $B(\bar C,r_0)$ and its complement. On the complement, the Gaussian factor is bounded by
\[
\exp\{-r_0^2/(4\sigma^2)\}\,\exp\{-\rho(C,\bar C)^2/(4\sigma^2)\},
\]
and Proposition~\ref{prop:geometric_foundation} makes the resulting integral exponentially small relative to any power of $\sigma$.

Inside $B(\bar C,r_0)$, use normal coordinates $v\in T_{\bar C}\Corr_p$. The squared distance is $\|v\|^2$, and the volume density has expansion
\[
\sqrt{\det g(v)}=1-\frac16\Ric_{ij}(\bar C)v^iv^j-\frac1{12}\nabla_k\Ric_{ij}(\bar C)v^iv^jv^k+O(\|v\|^4).
\]
The cubic term is odd and integrates to zero against the centered Gaussian kernel. Therefore
\[
Z=(2\pi\sigma^2)^{d/2}\left\{1-\frac16\E[\Ric_{ij}Y^iY^j]+O(\sigma^4)\right\},
\]
where $Y\sim N(0,\sigma^2I_d)$. Since $\E[Y^iY^j]=\sigma^2\delta_{ij}$, the correction is $-\Scal(\bar C)\sigma^2/6$. Uniformity over compact sets follows if the injectivity radius is bounded below and curvature derivatives are bounded on the corresponding compact tube. \qed

\subsection{Proof of Theorem~\ref{thm:sampler_correctness}}

The two samplers require separate arguments. Detailed balance establishes stationarity for Metropolis--Hastings, while the law of large numbers gives consistency for importance sampling and resampling.

For Algorithm~\ref{alg:mh_sampler}, the proposal density is symmetric. The acceptance ratio is the usual Metropolis--Hastings ratio for the unnormalized target density in \eqref{eq:coordinate_density}. Detailed balance follows, so the chart target is stationary. Because a Gaussian random walk has positive density on all of $\R^d$ and the target density is positive and continuous, the chain is irreducible and aperiodic. Standard Markov chain theory gives ergodic convergence for integrable functions.

For Algorithm~\ref{alg:sir_sampler}, domination gives $q(x)>0$ wherever the target is positive. The weight $w(x)$ satisfies
\[
\E_q\{w(X)\}=\int \exp\left\{-\frac{\rho(C(x),\bar C)^2}{2\sigma^2}\right\}J_{\QA}(x)dx=Z(\bar C,\sigma).
\]
The strong law of large numbers gives $\widehat Z\to Z$ almost surely and self-normalized importance estimates converge almost surely for integrable test functions. Conditional on the proposal sample, resampling produces an empirical distribution with probabilities proportional to the normalized weights; for fixed $B$, the unconditional resampled distribution converges weakly to the target as $M\to\infty$. \qed
 
\bibliography{references}

\end{document}